\documentclass[aps,prd,superscriptaddress,A4paper,preprintnumbers,nofootinbib]{revtex4}
\usepackage{amssymb,amsmath,graphicx,epsfig,bm,color}
\usepackage{bm}
\usepackage{float}
\usepackage{slashed}
\usepackage{verbatim} 
\newcommand{\be}{\begin{eqnarray}}
\newcommand{\ee}{\end{eqnarray}}
\usepackage[colorlinks]{hyperref}
\usepackage{leftidx}
\usepackage{booktabs}
\usepackage{latexsym}
\usepackage{revsymb}
\usepackage{cancel}
\usepackage{multirow}
\usepackage{caption}
\usepackage{xcolor}
\usepackage{subcaption}
\usepackage[normalem]{ulem}
\newcommand{\stkout}[1]{\ifmmode\text{\sout{\ensuremath{#1}}}\else\sout{#1}\fi}

\hypersetup{
    colorlinks=true,
    citecolor=blue,
    filecolor=magenta,      
    urlcolor=cyan,
}

\renewcommand{\d}{\mathrm{d}}

\newcommand{\xB }{x_{\scriptscriptstyle B}}
\newcommand{\sT}{{\scriptscriptstyle T}}

\definecolor{amol_color}{RGB}{200, 0, 200}
\definecolor{khatiza_color}{RGB}{191, 0, 0}
\definecolor{asmita_color}{rgb}{0.0, 0.1, 0.9}
\definecolor{rajesh_color}{rgb}{0.93, 0.53, 0.18}

\begin{document}

\title{Azimuthal asymmetries in $D$-meson and jet production at the EIC }

 \author{Khatiza Banu}
 \email{banu2765@iitb.ac.in}
 \affiliation{Department of Physics, Indian Institute of Technology Bombay, Mumbai-400076, India}
 \affiliation{Centre for Frontiers in Nuclear Science, Stony Brook University, Stony Brook, NY 11794-3800, USA}
 
 \author{Asmita Mukherjee}
 \email{asmita@phy.iitb.ac.in}
 \affiliation{Department of Physics, Indian Institute of Technology Bombay, Mumbai-400076, India}

\author{Amol Pawar}
 \email{194120018@iitb.ac.in}
 \affiliation{Department of Physics, Indian Institute of Technology Bombay, Mumbai-400076, India}
 
 \author{Sangem Rajesh}
\email{sangem.rajesh@vit.ac.in}
 \affiliation{Department of Physics, School of Advanced Sciences, Vellore Institute of Technology, Vellore,
Tamil Nadu 632014, India}

\date{\today}
\begin{abstract}
We study the azimuthal asymmetries in back-to-back leptoproduction of $D$-meson and jet to probe the gluon TMDs in an unpolarized and transversely polarized electron-proton collision at the kinematics of EIC. We give predictions for unpolarized cross-sections within the TMD factorization framework. 
In $D$-meson and jet formation, the only leading order contribution comes from the photon gluon fusion process.
We give numerical estimates of the upper bound on the azimuthal asymmetries with the saturation of positivity bounds; also, we present the asymmetries using a Gaussian parameterization of TMDs. We obtain sizable asymmetries in the kinematics that will be accessible at EIC.
\end{abstract}
\maketitle
 
\section{Introduction}\label{sec1}
 Transverse momentum dependent parton distribution functions (TMDs) have become the primary focus of research in hadron physics, as they encode the three-dimensional structure of hadron. TMDs depend on the parton's longitudinal momentum fraction ($x$) and its intrinsic transverse momentum ($\bm k_\perp$).  In contrast to the collinear parton distribution functions (PDFs) which can only provide one-dimensional tomography of the hadron, since they are dependent only on the parton's longitudinal momentum fraction; the TMDs give a three-dimensional momentum space description of the hadron in terms of its constituents.  TMDs are typically non-perturbative  in nature \cite{Mulders:1995dh}, and they can be  studied in processes like the semi-inclusive deep inelastic scattering process (SIDIS) \cite{Boer:1997nt, Bacchetta:2006tn} and Drell-Yan (DY) \cite{Tangerman:1994eh, Arnold:2008kf}.
In these processes, one observes a final hadron with transverse momentum  or a lepton pair which contains the footprint of the transverse momentum of the partons inside the proton. 
TMDs are not universal, since their operator definition  contains a gauge link (Wilson line),  making them process dependent \cite{Buffing:2013kca}. Unlike quark TMDs, which only need one gauge link to be defined in a  gauge-invariant way, the gluon TMD operators require two gauge links; which depend on the process being considered.  
 These gauge links could either be  future pointing gauge links (final state interactions)  or past-pointing gauge links (initial state interactions) or a mixture of both.
In small-$x$ physics, these two types of TMDs (known as unintegrated gluon distributions), are known  as the Weizs$\ddot a $cker-Williams (WW) gluon distribution \cite{McLerran:5909l, Kovchegov:1998bi} 
and the dipole gluon distribution \cite{Dominguez:8504fb}.
Both of these distributions have been commonly used in the literature and can be studied in different processes depending on the process-dependent gauge link structure.

At the leading twist, there are eight gluon TMDs. Among these, the Boer-Mulders function, $h_{1}^{\perp g}$, and the Sivers function, $f_{1T}^{\perp g}$  have gained a lot of attention in recent years.  Similar to this, we have quark TMDs, and the quark Sivers function is fairly well-known thanks to relentless experimental and theoretical  efforts \cite{Bury:2021sue, Boglione:2018dqd, Anselmino:2016uie}.  
However, little is known about the gluon TMDs.
The linearly polarized gluon distribution was first discussed in \cite{Mulders:2001pj} and calculated in a model in \cite{Meissner:2007}. The Boer-Mulders TMD represents the density of linearly polarized gluons inside an unpolarized proton. The $h_{1}^{\perp g}$ is a $T$ (time-reversal) - and chiral-even function, hence it is non-zero even in the absence of initial-state interactions (ISI) or final-state interactions (FSI) \cite{Mulders:2001pj}.
More information about the linearly polarized gluon TMDs can be obtained by calculating the $\cos2\phi_T$  type of azimuthal asymmetry, which is a ratio of linearly polarized gluon TMD to unpolarized gluon TMD.
The gluon Sivers function describes the distribution of unpolarized gluons inside a transversely polarized hadron. The correlation between the intrinsic transverse momentum of a parton and polarization of a proton leads to the asymmetric  distribution of final-state particles, which is the so-called Sivers asymmetry \cite{Sivers:1989cc, Sivers:1990fh}. Sivers asymmetry helps in understanding the spin crisis \cite{Ji:2002xn}. The first transverse moment of the Sivers function is related to the twist-three Qiu-Sterman function \cite{Qiu:1998ia, Qiu:1991pp}. The $f_{1T}^{\perp g}$, $T$-odd, changes sign in the SIDIS process compared to that in the DY process \cite{Collins:2002kn}. The ISI and FSI  play an important role in the Sivers asymmetry; in general, the gluon Sivers function (GSF) can be expressed in terms of  two independent GSFs that are called $f$-type and $d$-type GSF, respectively \cite{Brodsky:2002cx, Collins:2002kn, Belitsky:2002sm, Ji:2002aa, Boer:2003cm}. The $f$-type GSF contains $(++$ or $--$ ) gauge link and in the literature of small-$x$ physics called as  Weizsacker-Williams (WW) gluon distribution. The $d$-type GSF contains a $(+-)$  gauge link and is called dipole-type gluon distribution. 
The non-zero quark Sivers function has been extracted  from the HERMES \cite{Airapetian:1994, HERMES:1999ryv} and COMPASS \cite{COMPASS:2005csq, ADOLPH2012383} experiments, but the gluon Sivers function remains unknown, although initial attempts have been made \cite{DAlesio:2015fwo, DAlesio:2018rnv} to extract the GSF from RHIC data \cite{PHENIX:2013wle} in the mid-rapidity region.

Theoretical investigations indicate that the  gluon TMDs could be probed in the production of heavy-quark pair or dijet \cite{Boer:2011, Pisano:2013cya, Efremov:2017iwh}, $J/\psi$-photon \cite{Chakrabarti:2022rjr}, and $J/\psi$-jet \cite{DAlesio:2019qpk, Kishore:2022ddb, Kishore:2019fzb} at the Electron-Ion Collider (EIC), where  the transverse momentum imbalance of the pair is measured. 
Azimuthal asymmetries have been studied in various processes, including the production of $J/\psi$ \cite{Kishore:2018ugo, Kishore:2021vsm, DAlesio:2021yws, DAlesio:2020eqo, Rajesh:2018qks},  photon pair \cite{Qiu:2011}, and Higgs boson-jet \cite{Boer:2014lka}  production at LHC  have been proposed to probe the gluon TMDs. 
In these processes, the transverse momentum of the pair ($q_T$) is smaller than the individual transverse momentum ($K_{\perp}$) which allows us to use the TMD factorization. Transverse single-spin asymmetry (SSA) has been studied for inclusive $D$-meson production both in electron-proton \cite{GodboleRohiniMandKaushikPhysRevD.97.076001} and proton-proton \cite{ AnselminoMPhysRevD.70.074025, GodboleRohiniMPhysRevD.94.114022, Alesio:2017um} collision processes  within the generalized parton model framework. The SSA in the electroproduction of $D$-meson has been studied within the twist-three approach using the collinear factorization framework \cite{Kang:2008qh}.

In the present article, we present a calculation of azimuthal asymmetries in back-to-back electroproduction of a $D$-meson and jet in the process $e+p\rightarrow e+ D+\mathrm{jet}+X$ within a TMD factorization framework.
We consider the cases where the proton is unpolarized as well as transversely polarized. Our main focus is on calculating the azimuthal asymmetries such as $\cos2\phi_T$, $\cos2(\phi_T-\phi_{\perp})$ and $\sin(\phi_S-\phi_T)$. 
These asymmetries allow us to probe linearly polarized gluon TMD and Sivers TMD. 
In $D$-meson and jet production, at leading order (LO) in strong coupling constant ($\alpha_s$),  only the partonic channel of virtual photon-gluon fusion,  ${\gamma^\ast + g \to c + \bar{c}}$, contributes while the quark channel contributes at next-to-leading order (NLO). At LO, the produced charm quark fragments to form the $D$-meson and the anticharm quark evolves into the jet. The $D$-meson in the final state is the lightest meson containing a single charm quark (or antiquark). We consider the kinematics where the produced charm and anticharm quarks in the hard process have an almost equal magnitude of transverse momenta, but they are in opposite directions as shown in  Fig.~\ref{fig_angles}. 
The  produced $D$-meson (which we assume to be collinear to the fragmenting quark) and jet are almost back-to-back in the transverse plane. In this kinematical region, the total transverse momentum ($q_T$) of the system is much smaller than the individual transverse momentum ($K_\perp$) of the outgoing particles $i.e.~|\bm q_T| \ll |\bm K_\perp|$. Only in this region, the intrinsic transverse momentum can have significant effects, and we can assume that the TMD factorization is valid for the given process.

This paper is organized as follows: In Sec.\ref{sec2} we introduce the relevant kinematics of $D$-meson and jet production in the SIDIS process to calculate the azimuthal asymmetries. In Sec.\ref{sec3} we give the derivation to calculate the unpolarized scattering cross-section using TMD factorization. 
The azimuthal asymmetries that give  direct access to gluon TMDs are given in Sec.\ref{sec4} as well as the parameterization of the TMDs. 
In Sec.\ref{sec5} the numerical results and discussion are given. Finally, we conclude and an appendix is given at the end.

\section{Formalism}\label{sec2}
We start this section by specifying our notation and kinematics of SIDIS. We consider the production of a $D$-meson and a jet  in (un)polarized $ep$ scattering process, 
\be
e(l)+p^\uparrow(P)\rightarrow e(l^\prime)+ D(P_h)+\mathrm{jet}+X\,,
\ee
the $4$-momenta of each particle is given in the round brackets, and the transverse polarization of the proton is represented with an arrow in the superscript. 
For the collision energy that we are interested in this work, the process involves one-photon exchange. We define the virtual photon momentum, $q =l-l^{'}$, and its invariant mass as $Q^{2} = -q^{2}$.
We have considered the photon-proton center-of-mass (cm) frame, in which the photon and proton move along the $z$-axis.
We define the following kinematical variables,
\begin{eqnarray}
s=(l+P)^{2}, ~~~~~~~~~~~ \xB=\frac{Q^2}{2P\cdot q},~~~~~~~~~~~~ y=\frac{P\cdot q}{P\cdot l},
\end{eqnarray}
where $s$ is the square of the  energy of the electron-proton system in the cm frame, $Q^2$ is the virtuality of the photon, $\xB$ is known as the Bjorken variable and $y$ is called the inelasticity variable which is physically interpreted as the fraction of the electron energy transferred to the photon. These variables are related to each other through the relation $Q^2\approx\xB ys$.

We introduce two light-like vectors $n_+$ and, $n_-$ which obey the relations $n_+^2=n_-^2=0$ and $n_+\cdot n_-=1$.  The 4-momenta of the target 
system proton $P$ and  virtual photon $q$ can be written as,
\be\label{pq4m}
P^\mu&=&n^\mu_-+\frac{M_p^2}{2}n^\mu_+\approx n^\mu_-\,,\nonumber\\
q^\mu&=&-\xB n^\mu_-+\frac{Q^2}{2\xB}n^\mu_+\approx -\xB P^\mu+(P\cdot q)n^\mu_+\,,
\ee
with $P^2= 0$.
 The invariant mass of virtual photon-proton system is defined as $W^2_{\gamma p}=(q+P)^2$ and can also be expressed as $W^2_{\gamma p}=\frac{Q^2(1-\xB )}{\xB }=ys-Q^2$ and the mass of the proton is denoted by $M_p$. 
 We can express all the momenta in terms of $ n^\mu_-=P^\mu$ and
 $n^\mu_{+} = (q^\mu + \xB P^\mu)/P\cdot q $. 
 The 4-momentum of the incoming lepton reads as,
\be\label{q4m2}
l^\mu&=&\frac{1-y}{y}\xB n^\mu_-+\frac{1}{y}\frac{Q^2}{2\xB }n^\mu_++\frac{\sqrt{1-y}}{y}Q\hat{l}^\mu_\perp\,,
\ee
with $l^2=0$ and $\hat{l}_\perp^\mu$ is the unit transverse vector.\par 

The leading order (LO) partonic subprocess $ \gamma^*(q) +g(k) \rightarrow c(p_1) + \bar{c}(p_2) $ contributes to the process considered above.
\begin{figure}[H]
\begin{center} 
\includegraphics[height=5cm,width=10cm]{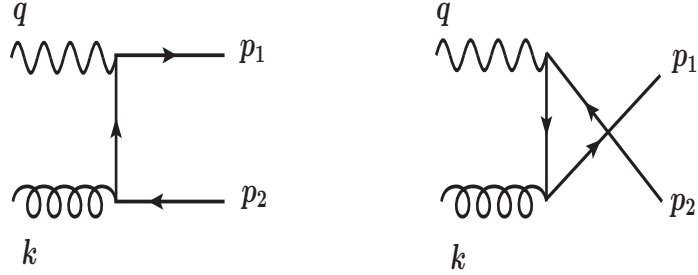}
\end{center}
\caption{\label{fig_feyn} Feynman diagrams for partonic subprocess $ \gamma^*(q) +g(k) \rightarrow c(p_1) + \bar{c}(p_2) $ at LO.}
\end{figure}
 In terms of light-like vectors, the four-momentum of the initial gluon is given as,
\be
k^\mu&\backsimeq&x P^\mu +k^\mu_{\perp g}, \,
\ee
where, $x$ and $k_{\perp g}$ are respectively the light-cone momentum fraction and the intrinsic transverse momentum of the incoming gluon with respect to the parent proton direction.
The four momenta of the produced heavy quarks in terms of light-like vectors are given as,  
\be\label{fv:pc}
p_1^\mu=z_1(P\cdot q)n_+^\mu + \frac{m_c^2+\bm {p}_{1\perp}^2} {2z_1 P\cdot q} P^\mu+ p_{1\perp}^\mu\,\nonumber \\
p_{2}^\mu=z_2(P\cdot q)n_+^\mu + \frac{m_c^2+\bm {p}_{2\perp}^2}{2z_2 P\cdot q} P^\mu+ p_{2\perp}^\mu\,,
\ee
where $z_1=\frac{P \cdot p_1}{P\cdot q}$ and $z_2=\frac{P\cdot p_2}{P\cdot q}$ are the momentum fractions of the charm and anti-charm quarks, and $m_c$ is the mass of the produced charm and anti-charm quark. The $p_{1 \perp}$ and $p_{2 \perp}$ are the transverse momenta of charm and anti-charm quarks, respectively.
The $4$-momentum of the $D$-meson in terms of light-like vectors can be written as,
\be
P_h^\mu=z_h(P\cdot q)n_+^\mu + \frac{m_h^2+\bm {P}_{hT}^2}{2z_h P\cdot q} P^\mu+ P_{hT}^\mu.\
\ee

The inelastic variable  $z_h=\frac{P\cdot P_h}{P\cdot q}$ is the energy fraction of the virtual photon taken by the observed $D$-meson in proton rest frame and $m_h$ is the mass of the $D$-meson. The 4-momentum of the charm quark, $p_1^\mu$, given in Eq.\eqref{fv:pc}, can be parametrized using the momentum fraction as,
\be\label{fv:pc2}
p_1^\mu=\frac{1}{z}P^\mu_h+ \frac{1}{2P\cdot P_h}\left(m_c^2 z- \frac{m_h^2}{z}\right)P^\mu,\,
\ee
where $z=\frac{P\cdot P_h}{P\cdot p_1}=\frac{z_h}{z_1}$ is the momentum fraction of $D$-meson in charm quark frame. Using Eq.\eqref{fv:pc} and \eqref{fv:pc2}, the transverse momentum of the charm quark and the fragmented $D$-meson are related by the following equation \cite{Koike:2011ns},
\be\label{eq:pcpd}
p_{1\perp}^\mu=\frac{1}{z} P_{hT}^\mu.
\ee
The Mandelstam variables are defined as,
\be
\hat{s}=(q+k)^2=-Q^2+2q.k, ~~~ \hat{u}=(k-p_1)^2=m_c^2-2k.p_1,~~~  
\hat{t}=(q-p_1)^2=m_c^2-Q^2-2q.p_1\,.
\ee
\section{Scattering cross-section}\label{sec3}

In the $ep$ scattering process, we consider the kinematical region in which the charm and anti-charm quarks are produced in a back-to-back configuration. In this kinematics, we use TMD factorization to write the cross-section. 
Here, the total transverse momentum of the system  ${q_T}$ with respect to the lepton plane is small compared to the virtuality of the photon $Q$ and to the mass of $D$-meson $m_h$.
The total differential scattering cross-section for the $e+p\to e+ c(p_1)+\bar{c}(p_2)+X$  the process can be written as,
\begin{equation}\label{eq:1}
 \begin{aligned}
 \d\sigma^{ep\to e+c\bar{c}+X}={}&\frac{1}{2s}\frac{\d^3{\bm l^\prime}}{(2\pi)^32E_{l^\prime}}
\frac{\d^3{\bm p}_1}{(2\pi)^32E_1}\frac{\d^3{\bm p}_{2}}{(2\pi)^32E_{2}}\int \d x \,\d^2 {\bm k}_{\perp g} \, \d z\,(2\pi)^4 \,\delta^4(q+k-p_1-p_2)\\
& \times\frac{1}{Q^4}L^{\mu\nu}(l,q)\,\Phi^{\rho\sigma}_g(x,{\bm k}_{\perp g})\, H_{\mu\rho}^{\gamma^\ast g \to c\bar{c}}H_{\nu\sigma}^{\ast;\gamma^\ast g \to c\bar{c}}\, D(z).
\end{aligned}
\end{equation}
Here, $E_i$ is the energy of the corresponding particle. In the $ep$ scattering process, $D$-meson is produced from the fragmentation of produced charm quark. In our kinematics where the $D$-meson and jet are in almost back-to-back configuration, we have neglected the intrinsic transverse momentum of the $D$-meson with respect to the charm quark in the hard part (this can be seen from Eq.(\ref{eq:pcpd}));  which is small compared to the large transverse momentum ${\bm P}_{hT}$.  In other words, we can consider the $D$ meson to be collinear to the fragmenting heavy quark. This gives the collinear fragmentation function $D(z)$, instead of the TMD fragmentation function in our expression. The differential scattering cross section for the process $e+p\to e+D(P_h)+\bar{c}(p_2)+X $ can be written as \cite{Pisano:2013cya},
\begin{equation}\label{eq:2d2}
 \begin{aligned}
 \d\sigma^{ep\to e+D+\bar{c}+X}={}&\frac{1}{2s}\frac{\d^3{\bm l^\prime}}{(2\pi)^32E_{l^\prime}}
\frac{\d^3{\bm P}_h}{(2\pi)^32E_h}\frac{\d^3{\bm p}_{2}}{(2\pi)^32E_{2}}\int \d x \,\d^2 {\bm k}_{\perp g} \, \d z \,(2\pi)^4 \,\delta^4(q+k-p_1-p_2)\\
& \times\frac{1}{Q^4}L^{\mu\nu}(l,q)\,\Phi^{\rho\sigma}_g(x,{\bm k}_{\perp g})\, H_{\mu\rho}^{\gamma^\ast g \to c\bar{c}}H_{\nu\sigma}^{\ast;\gamma^\ast g \to c\bar{c}}\, D(z)\,J(z)\,
\end{aligned}
\end{equation}
where $D(z)$ is the collinear fragmentation function describing the fragmentation of $D$-meson from the charm quark, it gives the number density of finding a  $D$-meson inside the charm quark with light-cone  momentum fraction $z$ in the charm quark frame.

The invariant phase space of the charm quark is related to the phase space of the final $D$-meson through the Jacobian factor $J$ as, 
\begin{eqnarray}
\frac{\d^3{\bm p}_1}{E_1}=J(z)\frac{\d ^3{\bm P}_h}{E_h}\,
~~~~~~~~{\rm with}~~~~~~~~
J=\frac{1}{z^3}\frac{E_h}{E_1}.
\end{eqnarray} \\
The momentum conservation delta function, given in  Eq.\eqref{eq:1},  can be decomposed as follows
\begin{equation}\label{dfun1}
\begin{aligned}
 \delta^{4}\bigl(q+k-p_1-p_2\bigr)
 & =\frac{2}{ys}\delta\bigl(1-z_1-z_2\bigr)\delta\left(x-\frac{z_2(m^2_c+ \bm{p}_{1\perp}^2)+z_1(m^2_c+\bm{p}^{2}_{2\perp}) + z_1 z_2Q^2}{z_1z_2ys}\right )\delta^{2}\bigl(\bm{k}_{\perp g}- \bm{p}_{1\perp}- \bm{p}_{2\perp}\bigr)\;,
\end{aligned}
\end{equation}
After substituting Eq.\eqref{eq:pcpd} in Eq.\eqref{dfun1} we get,
\begin{equation}\label{dfun2}
\begin{aligned}
 \delta^{4}\bigl(q+k-p_1-p_2\bigr)
 & =\frac{2}{ys}\delta\bigl(1-z_1-z_2\bigr)\delta\left(x-\frac{z_2(m^2_c+ \bm{P}_{hT}^2/z^2)+z_1(m^2_c+\bm{p}^{2}_{2\perp}) + z_1 z_2Q^2}{z_1z_2ys}\right )\delta^{2}\bigl(\bm{k}_{\perp g}- \frac{\bm{P}_{hT}}{z}- \bm{p}_{2\perp}\bigr)\,.
\end{aligned}
\end{equation}
The phase space of the outgoing particles is given by
\be\label{felectron}
 \frac{\d^3{\bm l^\prime}}{(2\pi)^32E_{l^\prime}}&=&\frac{1}{16\pi^2}\d Q^2\d y\,, \quad 
 \frac{\d^3{\bm P}_h}{(2\pi)^32E_h}=\frac{\d^2{\bm P}_{h T}\d z_{h}}{(2\pi)^3 2z_h}\,, \quad \frac{\d^3{\bm p}_2}{(2\pi)^32E_2}=\frac{\d^2{\bm p}_{2\perp}\d z_{2}}{(2\pi)^3 2z_2}.
\ee

We shift to the coordinate system which is more suitable for back-to-back scattering, for which we define the sum ($\bm q_T$) and difference of the transverse momenta ($\bm K_\perp$) of the outgoing quark and antiquark as,
\be\label{eq:newfv}
\bm q_T=\frac{\bm P_{hT}}{z}+\bm p_{2\perp}\,, \quad
\bm K_\perp= \frac{\frac{\bm P_{hT}}{z}-\bm p_{2\perp}}{2}\,.
\ee
Now the magnitude of the transverse momenta of the outgoing charm and anti-charm quark are almost equal. 
 In the back-to-back $D$-meson and jet production, the total transverse momentum, $\bm q_T$,  of the system, is much smaller than the individual transverse momentum of the outgoing particles $\bm K_\perp$ i.e., $|\bm q_T| \ll |\bm K_\perp|$. Using Eq.\eqref{eq:newfv}, we get 
 \be\label{eq:newfq}
\d^2{\bm  P}_{hT} \d^2 {\bm  p}_{2\perp}=z \d^2{\bm  q}_{T} \d^2 {\bm  K}_{\perp}.
\ee

 In Eq.\eqref{eq:2d2}, the leptonic tensor $L^{\mu\nu}$ has the standard form
\begin{eqnarray}\label{lep:ep}
L^{\mu\nu}&=&e^2 Q^2\left(-g^{\mu\nu}+\frac{2}{Q^2}(l^\mu l^{\prime\nu}+l^\nu l^{\prime\mu})\right)\nonumber\\
&=&e^2 Q^2\left(-g^{\mu\nu}+\frac{2}{Q^2}(2l^\mu l^{\nu}-l^\mu q^\nu- l^{\nu} q^\mu)\right)\,,
\label{eq:lep1}
\end{eqnarray}
where $e$ is the electronic charge, and we average over the spins of the initial lepton.  The 4-momentum of the final scattered lepton is $l^\prime=l-q$. By using Eq.\eqref{pq4m}, the leptonic tensor can be recast in the following form
\begin{equation}\label{lep:ep1}
 \begin{aligned}
L^{\mu\nu}={}&e^2 \frac{Q^2}{y^2}\Big[-(1+(1-y)^2)g_T^{\mu\nu}+4(1-y)\epsilon_{ L}^\mu \epsilon_{ L}^\nu+4(1-y)\left(\hat{l}^\mu_\perp \hat{l}^\nu_\perp +\frac{1}{2} g_T^{\mu\nu} \right)\\
&+2(2-y)\sqrt{1-y}\left(\epsilon_{ L}^\mu \hat{l}^\nu_\perp + \epsilon_{ L}^\nu \hat{l}^\mu_\perp \right)\Big]\,,
\end{aligned}
\end{equation}
where the transverse metric tensor is defined as $
g_{T}^{\mu\nu}=g^{\mu\nu}-n_+^{\mu}n_-^{\nu}-n_+^{\nu}n_-^{\mu}$, and  the light-like vectors can be written as below using Eq.\eqref{pq4m}
\be\label{ll_1}
n_-^\mu=P^\mu\,,~~~~~~~~n_+^\mu=\frac{1}{P\cdot q}(q^\mu+x_B P^\mu)\,.
\ee
The longitudinal polarization vector of the virtual photon is given as,
\be
\epsilon_{ L}^\mu(q) =\frac{1}{Q}\left(q^\mu +\frac{Q^2}{P\cdot q} P^\mu\right)\,,
\ee 
 with $\epsilon_{ L}^2(q)=1$ and $\epsilon_{ L}^\mu(q) q_\mu=0$. The factor $H$ in Eq.\eqref{eq:2d2} contains the scattering amplitude of $\gamma^\ast(q) +g(k) \rightarrow c + \bar{c}\,$  partonic process; the corresponding Feynman diagrams are shown in Fig.~\ref{fig_feyn}. 
\begin{figure}[H]
\begin{center} 
\includegraphics[height=5cm,width=10cm]{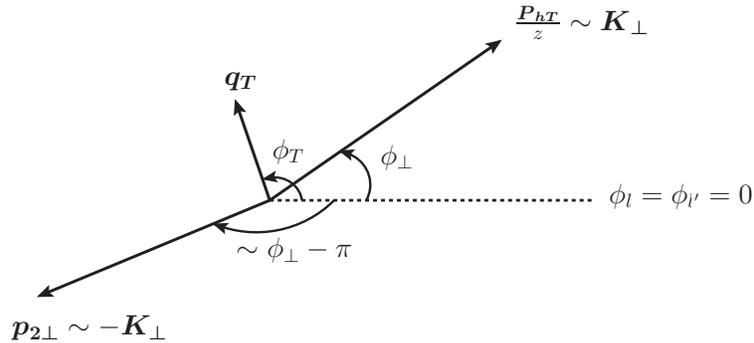}
\end{center}
\caption{\label{fig_angles} Representation of the azimuthal angles in {the production of} $D$ meson and jet in SIDIS process.}
\end{figure}

In Eq.\eqref{eq:2d2}, the gluon correlator $\Phi_g^{\mu\nu}$, {is a nonperturbative quantity that contains the dynamics of gluons inside a proton}. At leading twist, for an unpolarized proton, the gluon correlator parametrized in terms of two gluon TMDs as \cite{Mulders:2001pj},

\be\label{gc:un}
\Phi_{U}^{\mu\nu}(x,{\bm k}_{\perp g})=\frac{1}{2x}\Bigg\{-g_{T}^{\mu\nu}f_1^g(x,{\bm k}^2_{\perp g})+\left(\frac{k_{\perp g}^{\mu} k_{\perp g}^{\nu}}{M_p^2}+g_{T}^{\mu\nu}\frac{{\bm 
k}^2_{\perp g}}{2M_p^2}\right)h^{\perp g}_1(x,{\bm k}^2_{\perp g})\Bigg\}\,,
\ee
where $f_1^g$ and $h_1^{\perp g}$, $T$-even TMDs, encode the distribution of unpolarized and linearly polarized gluons for a given collinear momentum fraction $x$ and the transverse momentum $k_{\perp g}$, respectively. These TMDs can be non-zero even if initial and final state interactions are absent in the process. Similarly, for the transversely polarized proton, \cite{Mulders:2001pj} we have
\be
\begin{aligned}\label{gc:T}
\Phi_T^{\mu\nu}(x,{\bm k}_{\perp g} )={}& 
\frac{1}{2x}\bigg \{-g^{\mu\nu}_T
    \frac{ \epsilon^{\rho\sigma}_T k_{\perp g \rho} S_{T\sigma}}{M_p}
f_{1T}^{\perp\,g}(x, {\bm k}_{\perp g}^2) + i \epsilon_T^{\mu\nu}
    \frac{ k_{\perp g} \cdot  S_T}{M_p} g_{1T}^{g}(x, {\bm k}_{\perp g}^2) \\
&+  
\frac{k_{\perp g \rho}\epsilon_{T}^{\rho\{ \mu}k_{\perp g}^{\nu\}}} 
{2M_p^2} \frac{k_{\perp g }\cdot S_T}{M_p} h_{1T}^{\perp g}(x, {\bm k}_{\perp g}^2)
-\frac{k_{\perp g \rho}\epsilon_T^{\rho \{\mu}S_T^{\nu\}}+ 
S_{T\rho}\epsilon_T^{\rho\{\mu}k_{\perp g}^{\nu\}}}{4M_p}h_{1T}^{g}(x, {\bm k}_{\perp g}^2)  
 \bigg \}\,,
\end{aligned}
\ee
where the notations are: the antisymmetric tensor $\epsilon_T^{\mu\nu}=\epsilon^{\mu\nu\rho\sigma} P_\rho n_{+\sigma}$ with $\epsilon_T^{12}=+1$ and the symmetrization tensor $p_{\sT \rho}\epsilon_{T}^{\rho\{ \mu}p_{\sT}^{\nu\}}=p_{\sT \rho}(\epsilon_T^{\rho\mu}p_{\sT}^\nu+\epsilon_T^{\rho\nu}p_{\sT }^\mu)$. In Eq.\eqref{gc:T}, we have three $T$-odd TMDs: the Sivers function, $f_{1T}^{\perp g}$,  describes the density of unpolarized gluons, while $h_{1T}^{\perp g}$ and $h_{1T}^g$ are linearly polarized gluon densities of a transversely polarized proton. The  $T$-even TMD, {$g_{1T}^g$} is the distribution of circularly polarized gluons in a transversely polarized proton, which does not contribute here since it is in the antisymmetric part of the correlator. \par

After performing the integration over $z_2$, $x$ and ${\bm k}_{\perp g}$  in Eq.\eqref{eq:2d2}, we get
\begin{equation}\label{eq:ff3}
 \begin{aligned}
 \frac{\d\sigma^{ep\to e+D+\bar{c}+X}}{\d Q^2 \d y \d z_h \d^2{\bm  q}_{T} \d^2 {\bm  K}_{\perp} }={}&\frac{1}{ys^2}\frac{1}{16(2\pi)^4}
\int \d z\, D(z)\,\\
& \times\frac{1}{Q^4}L^{\mu\nu}(l,q)\,\Phi^{\rho\sigma}_g(x,{\bm q}_{T})\, H_{\mu\rho}^{\gamma^\ast g \to c\bar{c}}H_{\nu\sigma}^{\ast;\gamma^\ast g \to c\bar{c}}\, \,\frac{zJ(z)}{z_h(1-z_1)}\,.
\end{aligned}
\end{equation}

\section{Azimuthal asymmetries }\label{sec4}
In the kinematics wherein  the $D$ meson and the jet are back-to-back in the transverse plane (as discussed above), we can  write the cross-section as the sum of unpolarized and transversely polarized cross-sections as \cite{Pisano:2013cya},
\begin{equation}
 \frac{\d\sigma}{\d Q^2 \d y \d z_h \d^2{\bm  q}_{T} \d^2 {\bm  K}_{\perp} }  \equiv \d\sigma (\phi_S, \phi_\sT) =    \d\sigma^U(\phi_\sT,\phi_\perp)  +  \d\sigma^T (\phi_S, \phi_\sT)  \,.
\label{eq:cs}
\end{equation}

The cross-section for the unpolarized proton is written as the linear sum of $\cos\phi_T~\mathrm{and}~\cos\phi_\perp$ harmonics convoluted with the fragmentation function,
\begin{align}\label{eq:Un}
\d\sigma^{U} & =\mathcal{N}\int \d z\bigg[\bigl(\mathcal{A}_{0}+\mathcal{A}_{1} \cos\phi_{\perp}+\mathcal{A}_{2} \cos2\phi_{\perp}\bigr)f_{1}^{g}(x,\bm{q}_{\sT}^{2})+\bigl(\mathcal{B}_{0}  \cos2\phi_{\sT}+\mathcal{B}_{1}  \cos(2\phi_{\sT}-\phi_{\perp})\nonumber\\
 & \qquad\quad+\mathcal{B}_{2}  \cos2(\phi_{\sT}-\phi_{\perp})+\mathcal{B}_{3}  \cos(2\phi_{\sT}-3\phi_{\perp})+\mathcal{B}_{4}  \cos(2\phi_{\sT}-4\phi_{\perp})\bigr)\frac{ \bm q_\sT^2  }{M_p^2} \,h_{1}^{\perp\, g} (x,\bm{q}_{\sT}^2)\bigg]\, D(z)\,.
\end{align}
{where $\mathcal{N}$ is the normalization factor given as,
\begin{equation}
\mathcal{N}=\frac{\alpha^2 \alpha_s e_c^2}{\pi y^3 s^2 x}. 
\end{equation}}
The coefficients mentioned in the above equation are the result of the contribution from different helicities of the virtual photon and the linearly polarized gluon.
For instance, if the azimuthal angle of the final scattered lepton is not measured, then only one modulation term in Eq.\eqref{eq:Un} is defined, and the cross-section is expressed as,
\begin{align}\label{eq:Un1}
\d\sigma^{U} & =\mathcal{N}\int \d z\bigg[\mathcal{A}_{0}f_{1}^{g}(x,\bm{q}_{\sT}^{2})+\mathcal{B}_{2}  \cos2(\phi_{\sT}-\phi_{\perp})\frac{ \bm q_\sT^2  }{M_p^2} \,h_{1}^{\perp\, g} (x,\bm{q}_{\sT}^2)\bigg]\,D(z)\,,
\end{align}
while in the case of a transversely polarized proton,
\begin{align}\label{eq:Tr}
\d\sigma^T
  & =  \mathcal{N}{\vert \bm S_\sT\vert}\,\int \d z \bigg[\sin(\phi_S -\phi_\sT) \bigl( \mathcal{A} _0 + \mathcal{A} _1 \cos \phi_\perp  + \mathcal{A} _2 \cos 2 \phi_\perp \bigr ) \frac{ |\bm q_\sT | }{M_p}f_{1T}^{\perp\, g} (x,\bm{q}_{\sT}^2)\nonumber \\
 & + \cos(\phi_S-\phi_\sT) \bigl( \mathcal{B} _0 \sin 2 \phi_\sT  +  \mathcal{B} _1 \sin(2\phi_\sT-\phi_\perp) + \mathcal{B} _2 \sin 2 (\phi_\sT-\phi_\perp) \nonumber \\
 &\quad+ \mathcal{B} _3 \sin( 2\phi_\sT-3\phi_\perp)  + \mathcal{B} _4\sin(2\phi_\sT- 4\phi_\perp)  \bigr) \frac{ |\bm q_\sT |^3  }{M_p^3} \,h_{1T}^{\perp\, g} (x,\bm{q}_{\sT}^2) \nonumber\\
& + \bigl (\mathcal{B}_0  \sin(\phi_S+\phi_\sT) + \mathcal{B}_1  \sin(\phi_S + \phi_T-\phi_\perp) + \mathcal{B}_2  \sin(\phi_S+\phi_\sT-2\phi_\perp) \nonumber \\
&  \quad  +  \mathcal{B}_3  \sin(\phi_S+\phi_\sT-3\phi_\perp) + \mathcal{B}_4  \sin(\phi_S+\phi_\sT-4 \phi_\perp) \bigr)\frac{ |\bm q_\sT | }{M_p} h_{1T}^{g} (x,\bm{q}_{\sT}^2) \bigg ]\, D(z)\,,
\end{align}
where  $\phi_S$, $\phi_\sT$ and $\phi_\perp$  are  the azimuthal angles of the three-vectors $\bm S_\sT$,  $\bm q_\sT$ and  $\bm K_\perp$ respectively, measured w.r.t. the lepton plane      
 ($\phi_{\ell}=\phi_{\ell^\prime}=0)$ as shown in Fig.~\ref{fig_angles}. The coefficients of the different angular modulations $\mathcal{A}_i$ with $i=0,1,2$ and $\mathcal{B}_j$ with $j=0,1,2,3,4$  are given in the Appendix. \ref{appen}.\par
The weighted azimuthal asymmetry gives the ratio of the specific gluon TMD over unpolarized $f_1^g$ and is defined as \cite{DAlesio:2019qpk},
\begin{align}
A^{W(\phi_S,\phi_\sT)} & \equiv 2\,\frac {\int\d \phi_S\, \d \phi_\sT \,\d\phi_\perp\, W(\phi_S,\phi_\sT)\,\d\sigma(\phi_S,\,\phi_\sT,\,\phi_\perp)}{\int \d \phi_S\,\d\phi_\sT \,\d\phi_\perp\,\d\sigma(\phi_S,\phi_\sT,\phi_\perp)} \,,
\label{eq:mom}
\end{align}
where the denominator is given by
\begin{align}\label{eq:f1}
\int \d\phi_S\,\d \phi_\sT \,\d\phi_\perp\,\d\sigma(\phi_S,\phi_\sT,\phi_\perp)  & =\int \d\phi_S\,\d \phi_\sT \,\d\phi_\perp\,\d\sigma^U(\phi_\sT,\phi_\perp)=  (2\pi)^3 \mathcal{N}\int \d z {\cal A}_0 f_{1}^{g}(x,\bm{q}_{\sT}^{2})\,D(z)\;.
\end{align}
By integrating over the azimuthal angle $\phi_\perp$, the transversely polarized cross-section, Eq.~(\ref{eq:Tr}), can be simplified further as,
\begin{align}\label{eq:sivers}
\int\d\phi_\perp\d\sigma^T & =  2\pi {\vert \bm S_\sT\vert}\, \frac{ |\bm q_\sT | }{M_p}\int  \d z\left [\mathcal{A} _0 \sin(\phi_S -\phi_\sT)    f_{1T}^{\perp\, g} (x,\bm{q}_{\sT}^2) -\frac{1}{2} \mathcal{B} _0\sin(\phi_S-3\phi_\sT)   \frac{ |\bm q_\sT |^2 }{M_p^2} \,h_{1T}^{\perp\, g} (x,\bm{q}_{\sT}^2) \right . \nonumber\\*
  &\qquad+\mathcal{B}_0  \sin(\phi_S+\phi_\sT)  h_{1}^{g} (x,\bm{q}_{\sT}^2) \bigg ]\,D(z)\,,
\end{align}
where we have used the relation 
\begin{equation}
h_1^g \equiv h_{1T}^g +\frac{\bm p_\sT^2}{2 M_p^2}\,  h_{1T}^{\perp\,g}\,
\label{eq:h1}
\end{equation}
where $h_1^g$ ($T$-odd), is the helicity flip gluon distribution which is chiral-even and vanishes upon integration of transverse momentum \cite{Boer:2016fqd}. In contrast, the quark distribution is chiral-odd {($T$-even)} and survives even after the transverse momentum integration.
The $h_1^{\perp\, g}$ gluon TMD could be extracted by studying the following two azimuthal asymmetries,
\begin{align}
A^{\cos 2 \phi_\sT}&= \frac{\bm q_\sT^2}{M_p^2}
\,\frac{\int \d z\,  {\cal B}_0 \,D(z)\, h_1^{\perp\, g}(x,\bm q_\sT^2 )}{\int \d z \, {\cal A}_0\, D(z)\, f_1^{g}(x,\bm q_\sT^2 )}\,,
\label{eq:cos2phiT}
\end{align}
and
 \begin{align}
A^{\cos 2 (\phi_\sT-\phi_\perp)}&= \frac{\bm q_\sT^2}{M_p^2}\, \frac{\int \d z\,  {\cal B}_2\, D(z)\,h_1^{\perp\, g}(x,\bm q_\sT^2 )}{\int \d z\,  {\cal A}_0\, D(z)\, f_1^{g}(x,\bm q_\sT^2 )}\,.
\label{eq:cos2phiT2phiP}
\end{align}
Using Eq.\eqref{eq:sivers} with ${\vert \bm S_\sT\vert}=1$ , one could {utilize the following} asymmetries to extract the $f_{1T}^{\perp g}$, $h_1^g$ and $h_{1T}^{\perp g}$ TMDs,
\begin{align}\label{eq:f1t_sivers}
A^{\sin(\phi_S-\phi_\sT)} & =  \frac{\vert \bm q_\sT\vert}{M_p}\, \frac{\int \d z\,  {\cal A}_0\, D(z)\, f_{1T}^{\perp\,g}(x,\bm q_\sT^2) }{\int \d z\,  {\cal A}_0\, D(z)\,f_1^g(x,\bm q_\sT^2)}\,,
\end{align}
\begin{align} \label{eq:h1g-asy}
A^{\sin(\phi_S+\phi_\sT)}  & =  \frac{\vert \bm q_\sT\vert}{M_p}\, \frac{\int \d z\,  {\cal B}_0\, D(z)\, h_{1 }^{g}(x,\bm q_\sT^2) }{\int \d z\,  {\cal A}_0\, D(z)\,f_1^g(x,\bm q_\sT^2)}\,,
 \end{align}
 and
\begin{align}\label{eq:h1Tg-asy}
A^{\sin(\phi_S-3\phi_\sT)}  & =   -  \frac{\vert \bm q_\sT\vert ^3}{2M_p^3}\,  \frac{\int \d z\,  {\cal B}_0\, D(z)\, h_{1T}^{\perp\,g}(x,\bm q_\sT^2)}{\int \d z\,  {\cal A}_0\, D(z)\, f_1^g(x,\bm q_\sT^2)} \,.
\end{align}

\subsection{Positivity Bounds}
 The upper limit of the azimuthal asymmetries as defined above, can be reached when the polarized gluon TMDs saturate the positivity bounds that are independent of any specific model \cite{Bacchetta:1999kz, Mulders:2001pj}. 
\begin{align}
\frac{\vert \bm q_\sT \vert }{M_p}\, \vert f_{1T}^{\perp \,g}(x,\bm q_\sT^2) \vert & \le   f_1^g(x,\bm q_\sT^2)\,,\nonumber \\
\frac{ \bm q^2_\sT }{2 M_p^2}\, \vert h_{1}^{\perp\,g}(x,\bm q_\sT^2) \vert & \le   f_1^g(x,\bm q_\sT^2)\,,\nonumber \\
\frac{\vert \bm q_\sT \vert }{M_p}\, \vert h_{1}^g(x,\bm q_\sT^2) \vert & \le   f_1^g(x,\bm q_\sT^2)\,,\nonumber \\
\frac{\vert \bm q_\sT \vert^3}{2 M_p^3}\, \vert h_{1T}^{\perp \,g}(x,\bm q_\sT^2) \vert & \le   f_1^g(x,\bm q_\sT^2)\,.
\label{eq:bounds}
\end{align}
 Using the positivity bounds on the gluon TMDs given in Eqs.\eqref{eq:cos2phiT}-\eqref{eq:h1Tg-asy}  and  for the fixed kinematical variables,  we obtain the following upper bounds on the absolute value of the $A^{\cos2\phi_T}$ and  $A^{\cos2(\phi_T-\phi_\perp)}$ asymmetries, 
\begin{align}\label{eq:ub}
\vert A^{\cos 2 \phi_\sT}\vert \leq 2\frac{ |{\cal B}_0|  }{{\cal A}_0}\,, \quad  \vert A^{\cos 2 (\phi_\sT-\phi_\perp)}\vert \leq 2\frac{ |{\cal B}_2|  }{{\cal A}_0}\,,
\end{align}
and the upper bound for the Sivers asymmetry, $A^{\sin(\phi_S-\phi_\sT)}$, becomes equal to one, while the upper bounds for the other asymmetries can be determined using their relations with other asymmetries, such as,
\begin{align}\label{eq:ub1}
\vert A^{\sin(\phi_S+\phi_\sT)}\vert = \frac12 \,\vert A^{\cos 2 \phi_\sT}\vert \,, \quad  \vert A^{\sin(\phi_S-3\phi_\sT)}\vert =\frac12\,\vert A^{\cos 2 (\phi_\sT-\phi_\perp)}\vert\,.
\end{align}
\subsection{ Gaussian parameterization of TMDs}
The numerical estimate of the asymmetries depends on the parameterization used for the TMDs. In this work, we estimate the asymmetries using Gaussian parameterization. 
For the unpolarized gluon TMD, we adopt a parameterization given by,
\begin{eqnarray}\label{eq:gauss_f1}
f_1^g(x,\bm{q}_\sT^2)=f_1^g(x,\mu)\frac{e^{-\bm{q}_\sT^2/\langle q_\sT^2\rangle}}{\pi\langle q_\sT^2\rangle}\,,
\end{eqnarray}
where $f_1^g(x,\mu)$ is the collinear gluon PDF at the probing scale $\mu=\sqrt{m^2_h+Q^2}$ \cite{Kniehl:2006mw}. We use MSTW2008 set \cite{martin2009parton} for collinear PDF. The Gaussian parameterization of TMDs with a Gaussian width  $\langle q_\sT^2\rangle=1~\mathrm{GeV}^2$ for gluons \cite{Alesio:2017um}.
We adopt the following Gaussian parameterization for the linearly polarized gluon TMD $h_1^{\perp g}$ as given in Ref. \cite{Boer:2011kf, Boer:2012bt}, 
\begin{eqnarray}
			\label{eq:gauss_h1p}
			h_1^{\perp g}(x,\bm{q}_\sT^2)=\frac{M_p^2f_1^g(x,\mu)}{\pi\langle q_\sT^2\rangle^2}\frac{2(1-r)}{r}e^{1-\frac{\bm{q}_\sT^2}{r\langle q_\sT^2\rangle}},
\end{eqnarray}
where $M_p$ is the proton mass, $r$ (with $0<r<1$), and the average intrinsic transverse momentum width of the incoming gluon, $\langle q_\sT^2\rangle$, are parameters of this model. In our numerical estimation, we take $r=1/3$ and $\langle q_\sT^2\rangle=1~\text{GeV}^2$.  \par
Similarly, for the gluon Sivers function (GSF) $f_{1T}^{\perp g}$, we have used the parameterization given in Ref. \cite{Bacchetta:2004jz,DAlesio:2018rnv,Anselmino:2005ea},
\begin{eqnarray}\label{eq:Sivers:par}
\Delta^{N} f_{g / p^{\uparrow}}\left(x, q_\sT\right)= \left(-\frac{2|\bm{q}_\sT|}{M_P}\right)f_{1T}^{\perp g}\left(x, q_\sT\right)=2\frac{\sqrt{2 e}}{\pi} \mathcal{N}_{g}\left(x\right) f_{g / p}\left(x\right)  \sqrt{\frac{1-\rho}{\rho}} q_\sT \frac{e^{-\bm{q}_\sT^2 / \rho\left\langle q_\sT^2\right\rangle}}{\left\langle q_\sT^2\right\rangle^{3 / 2}}\,,
\end{eqnarray}

with $0<\rho<1$. The $x$- dependence of the gluon Sivers function is encoded in the $\mathcal{N}_{g}\left(x\right)$ and it is generally written as,
\begin{eqnarray}\label{eq:Sivers:Ng}
\mathcal{N}_{g}\left(x\right)=N_{g} x^{\alpha}\left(1-x\right)^{\beta} \frac{(\alpha+\beta)^{(\alpha+\beta)}}{\alpha^{\alpha} \beta^{\beta}}.\,
\end{eqnarray}
The parameters $N_g$, $\alpha$, and $\beta$ are determined from global fits to experimental data on single spin asymmetries (SSAs) in inclusive hadron production processes \cite{DAlesio:2018rnv}, while the extracted best-fit parameters at $\left\langle q_\sT^2\right\rangle=1~\text{GeV}^2$  are 
\be
 N_g=0.25\,,\quad \alpha=0.6\,, \quad \beta=0.6\,,\quad \rho=0.1 \,.
\ee
\subsection{Fragmentation function of the $D$ -meson}

At leading order (LO) the charm quark produced in the virtual photon-gluon fusion process fragments to form the $D$ meson in the final state. In our kinematics, we can consider the $D$ meson to be collinear to the fragmenting heavy quark.  This means that the transverse momentum of the $D$-meson is related to the charm quark's transverse momentum through Eq.(\ref{eq:pcpd}).
The LO fragmentation function for the  $c \rightarrow D^{0}$ process is parameterized as,
\begin{equation}
    D(z,\mu_0)=\frac{N z(1-z)^2}{\left[(1-z)^2 + \epsilon z \right]^{2}}
\end{equation}
which is given by \cite{Kniehl:2006mw}. The parameters are $N= 0.694$, $\epsilon=0.101$, and are fitted using OPAL Collaboration data at CERN LEP-I at the $\mu_0=m_c=1.5$ GeV. The scale evolution of the collinear fragmentation function is given by the DGLAP equation. Here, we ignore the scale evolution of the fragmentation function. A similar approach is followed in \cite{ Alesio:2017um,AnselminoMPhysRevD.70.074025,GodboleRohiniMPhysRevD.94.114022,GodboleRohiniMandKaushikPhysRevD.97.076001}.
\section{Results and Discussions}\label{sec5}
\subsection{Unpolarized cross-section}
In this section, we present numerical results for  the unpolarized cross-section of $D$-meson and jet production in the SIDIS process. 
The LO contribution comes only from the gluon-initiated partonic subprocess, $i.e.$ $\gamma^\ast + g  \rightarrow c + \bar{c} \rightarrow D+\bar{c}$, whereas the contribution from the quark-initiated process {occurs} at NLO. After integrating over the azimuthal angles,  only the $\mathcal{A}_{0}$ term contributes to the unpolarized cross-section given in Eq.\eqref{eq:Un} and its expression is given in Appendix \ref{appen}. We used the Gaussian parameterization, given in Eq.\eqref{eq:gauss_f1} for the unpolarized transverse momentum dependent (TMD) gluon distribution function $f_{1}^{g}(x,\bm{q}_T^2)$.
We consider the situation, in which the produced $D$-meson and jet are almost back-to-back, with $q_T^2 \ll Q^2$ and $|\bm{q}_T| \ll |\bm{K}_\perp|$, which allows us to assume the TMD factorization for the cross-section. We estimate the cross-section at the cm energy of the EIC with $\sqrt{s}=140$ and $45$ $ \mathrm{GeV}$, and we choose the following kinematical constraints. The range of integration of the virtuality of the photon $(Q^2)$ is  $3<Q^2<100$ $\mathrm{GeV^{2}}$,
 the momentum fraction $z$ carried by the $D$-meson from the charm quark is in the range  $0<z<1$. The 
 inelasticity variable $y$ is fixed from the definition of the invariant mass of photon-proton system, denoted as $W_{\gamma p}$ and it is in the range $20<W_{\gamma p}<80$ $\mathrm{GeV}$ for $\sqrt{s}=140$ $\mathrm{GeV}$, and $10<W_{\gamma p}<40$ $\mathrm{GeV}$ for $\sqrt{s}=45$ $\mathrm{GeV}$. In this kinematics, $q_T$ is the sum of the transverse momenta of the outgoing charm and anti-charm quarks, which is equal to the transverse momentum of the initial gluon, $q_T$ varies in the range  $0<q_T<1$ $\mathrm{GeV}$. The transverse momentum of the outgoing particles, $i.e.$, the $D$-meson, and the jet, denoted as $K_\perp$, is considered to be greater than $2$ GeV.  This condition, $|\bm{q}_T| \ll |\bm{K}_\perp|$, implies that the $D$-meson and jet are produced almost back-to-back in the process. We have set the upper and lower bound on the momentum fraction of the hadron as $0.1<z_h <0.9$. 
To avoid the unphysical contribution from the endpoints of the $z_h$, we imposed the aforementioned kinematic restriction  on the $z_h$.
\par

\begin{figure}[H]
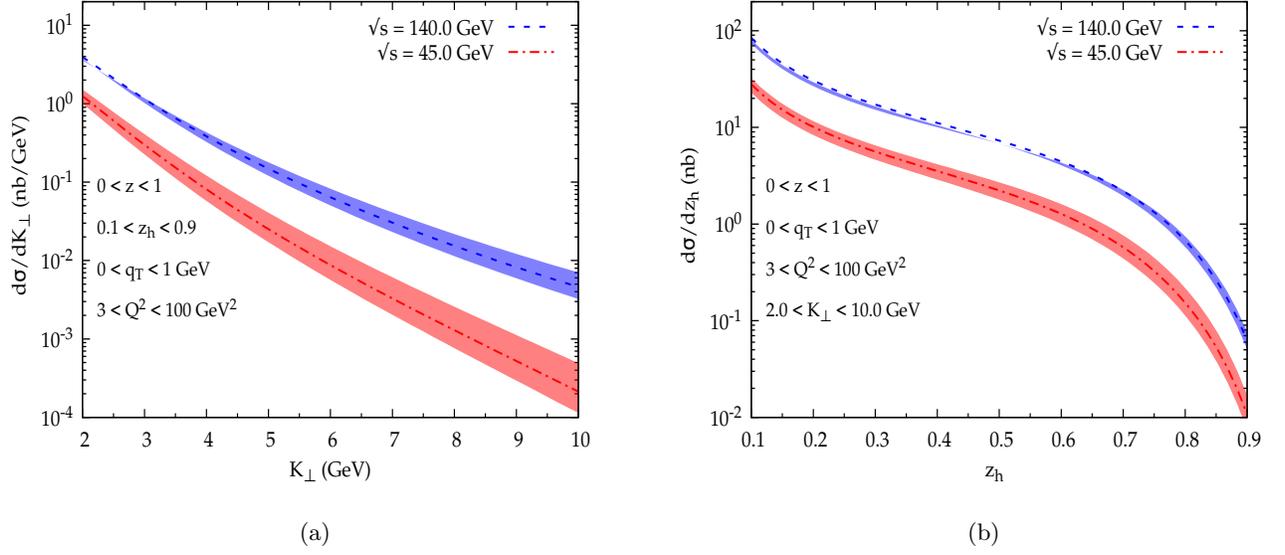

	\begin{center} 
		\begin{subfigure}{0.49\textwidth}
		\includegraphics[height=7cm,width=8.5cm]{Fig3a.pdf}
			\caption{}
		\end{subfigure}
	    \begin{subfigure}{0.49\textwidth}
	    \includegraphics[height=7cm,width=8.5cm]{Fig3b.pdf}
	        \caption{}
	    \end{subfigure}
	\end{center}
\caption{\label{fig:unc_scale} Unpolarized differential scattering cross-section of $e+p\rightarrow e +D+\mathrm{jet} +X$ process as a function of $K_\perp$ (left) and $z_h$ (right). For $K_{\perp}$ and $z_h$ variations, the $z$, $Q^2$ and $q_T$ are integrated over the regions $0<z<1$,~~ $3<Q^2<100~~\mathrm{GeV}^2$ and $0.0<q_T<1.0$ GeV. For $\sqrt{s}=140$ GeV, the range of $W_{\gamma p}$ is  $20<W_{\gamma p}<80$ GeV, 
while for $\sqrt{s}=45$ GeV, the range is  $10<W_{\gamma p}<40$ GeV.
For $K_{\perp}$ variation, we have taken $0.1<z_h<0.9$ and for $z_h$ variation,  $2.0<K_{\perp}<10.0$ GeV. 
The bands are obtained by varying the factorization scale in the range $\frac{1}{2} \mu < \mu < 2\mu $.}

\end{figure}

In Fig.~\ref{fig:unc_scale}, the unpolarized differential cross-section {is shown} as a function of the transverse momentum, $K_\perp$, of the $D$-meson and $z_{h}$, the momentum fraction carried by the $D$-meson from the virtual photon. The blue dashed line represents the cross-section for $\sqrt{s}=140$ $\mathrm{GeV}$, while the red dash-dotted line represents the cross-section for $\sqrt{s}=45$ $\mathrm{GeV}$. The cross-section is larger for higher cm energy due to the low momentum fraction $x$ region being probed at higher cm  energy compared to lower cm energy, and the density of gluons is higher in low $x$ region. In the left panel of Fig.~\ref{fig:unc_scale}, the cross-section falls  rapidly with increasing $K_\perp$ for lower cm energy, which is expected, as the production of high transverse momentum particles becomes less probable at lower energies.
 In the right panel of Fig.~\ref{fig:unc_scale} the scattering cross-section is plotted as a function of $z_h$,  which is obtained by integrating $K_\perp$ over the range $2< K_{\perp}<10~\mathrm{GeV}$. It is observed that the cross-section decreases as $z_h$ increases.  In Fig.~\ref{fig:unc_scale}, the band represents the theoretical uncertainty which is obtained by varying the factorization scale $\mu=\sqrt{Q^2+m_h^2}$ from $0.5 \mu $ to $2\mu $. The width of the uncertainty band in Fig.~\ref{fig:unc_scale} for $K_\perp$ variation becomes wider at high $K_\perp$, while it is narrow at small $K_\perp$. The scale uncertainty is expected to decrease at higher order in QCD.

\subsection{Upper bounds}
\begin{figure}[H]
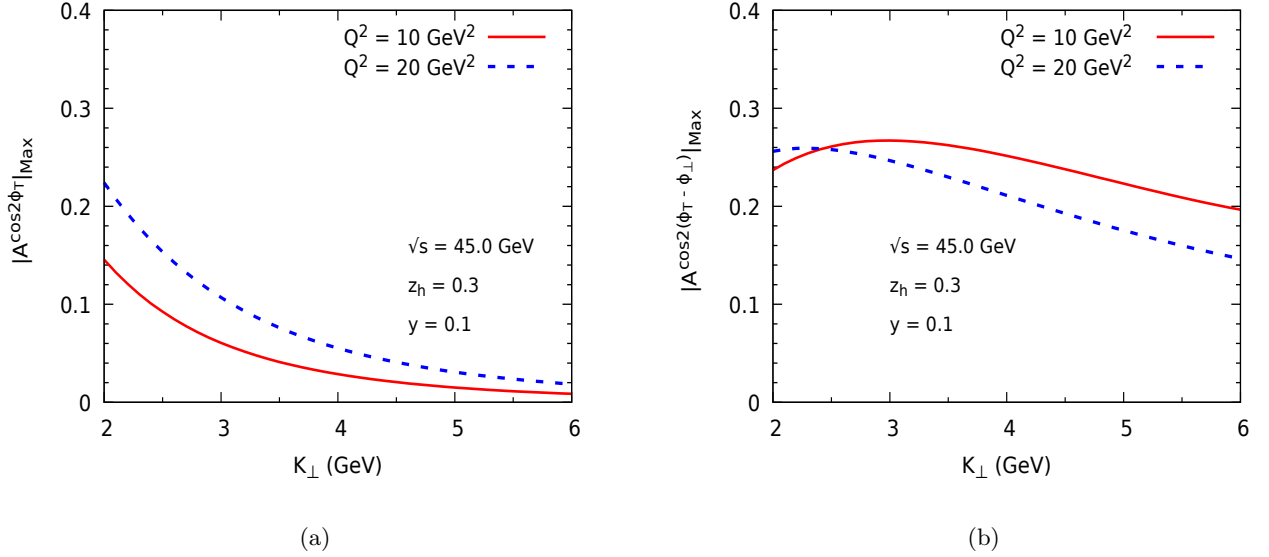

	\begin{center} 
		\begin{subfigure}{0.49\textwidth}
		\includegraphics[height=7cm,width=8.5cm]{Fig4a.pdf}
			\caption{}
		\end{subfigure}
	    \begin{subfigure}{0.49\textwidth}
	    \includegraphics[height=7cm,width=8.5cm]{Fig4b.pdf}
	        \caption{}
	    \end{subfigure}
     \caption{\label{fig:ub_vKp} Upper bound for the $A^{\cos2\phi_T}$ (left panel) and $A^{\cos2(\phi_{T}-\phi_\perp)}$ (right panel) azimuthal asymmetries in $e+p\rightarrow e +D+\mathrm{jet}+X$ process as function of $K_\perp$ at EIC $\sqrt{s}=45$ GeV for fixed values of $y=0.1$, $z_h=0.3$  and $Q^2=10$, 20 GeV$^2$. The kinematical variables $z$ and $q_T$ are integrated from [0,1].}
     \end{center}
\end{figure}

\begin{figure}[H]
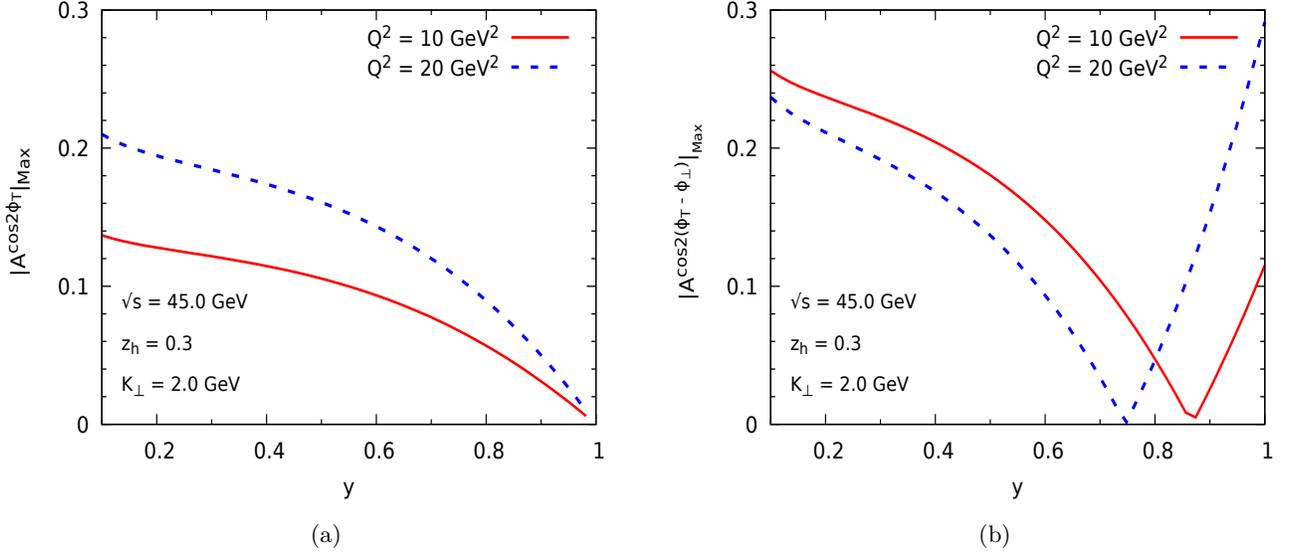

	\begin{center} 
		\begin{subfigure}{0.49\textwidth}
		\includegraphics[height=7cm,width=8.5cm]{Fig6a.pdf}
			\caption{}
		\end{subfigure}
	    \begin{subfigure}{0.49\textwidth}
	    \includegraphics[height=7cm,width=8.5cm]{Fig6b.pdf}
	        \caption{}
	    \end{subfigure}
     \end{center}
\caption{\label{fig:ub_vy} Upper bound for the $A^{\cos2\phi_T}$ (left panel) and $A^{\cos2(\phi_{T}-\phi_\perp)}$ (right panel) azimuthal asymmetries in $e+p\rightarrow e +D+\mathrm{jet}+X$ process as function of $y$ at {EIC $\sqrt{s}=45$ GeV for} fixed values of $z_h=0.3$, $K_\perp=2$ GeV  {and $Q^2=10$, 20 GeV$^2$}. The {kinematical variables} $z$ {and $q_T$ are integrated from [0,1]}.}

\end{figure}

In this section, we present the numerical estimates of the upper bounds for  $\cos2\phi_T$ and $\cos2(\phi_T-\phi_{\perp})$ asymmetries by saturating the positivity relations of the TMDs. 
In  Figs.~\ref{fig:ub_vKp}-\ref{fig:ub_vy}, we have plotted the upper bounds for the azimuthal asymmetries ${\cos2\phi_T}$ (left panel) and ${\cos2(\phi_{T}-\phi_\perp)}$ (right panel) in the process $e+p\rightarrow e +D+\mathrm{jet}+X$. The upper bound of the asymmetries depends on $\sqrt{s}$; with other kinematical variables fixed, we observed that the upper bound is about $3-4$\% higher for $\sqrt{s} =45$ GeV compared to $\sqrt{s} =140$ GeV. In the plot, we show the upper bound for $\sqrt{s} =45$ GeV.
We plotted the upper bound as a function of the transverse momentum, $K_\perp$, momentum fraction, $z_h$ and rapidity, $y$ at two different virtualities of the photon $Q^2=10,~20$ $\mathrm{GeV^{2}}$. We integrated the variables $z$ and $q_T$ variables within the range $0$ to $1$. The variation of $K_\perp$ is shown in Fig.~\ref{fig:ub_vKp} for fixed values of $z_h$ and $y$, the variation of $z_h$ is shown in Fig.~\ref{fig:ub_vzh} for the fixed values of $K_\perp$ and $y$, and the $y$ variation is shown in Fig.~\ref{fig:ub_vy} for the fixed values of $K_\perp$ and $z_h$.  

From Figs.~\ref{fig:ub_vKp}-\ref{fig:ub_vy}, it can be observed that the upper bound of $\cos2\phi_T$ azimuthal asymmetry increases with increasing virtuality of the photon $(Q^2)$. In Fig.~\ref{fig:ub_vKp}, one can see that for a given $Q^2$ the magnitude of the upper bound of $\cos2\phi_T$ azimuthal asymmetry decreases with increasing $K_\perp$. This can be attributed to the increase in the longitudinal momentum fraction of the initial gluon $x$ as $K_\perp$ increases, which leads to the vanishing of the gluon PDF as $x$ approaches $1$. 
The behavior of $K_\perp$ variation of the upper bound of $\cos2(\phi_T-\phi_{\perp})$ azimuthal asymmetry for two different resolutions of the photon exhibits a somewhat different  behavior, in small $K_\perp$ region, the high virtuality asymmetry dominates, while at  high $K_\perp$ the low virtuality curve dominates. With increasing $K_\perp$, the $\cos2(\phi_T-\phi_{\perp})$ azimuthal asymmetry initially increases, reaches a peak at around $2.5 ~\mathrm{GeV}$ for $Q^2=10~\mathrm{GeV}^2$ and $3 ~\mathrm{GeV}$ for $Q^2=20 ~\mathrm{GeV}^2$ and then decreases. Qualitatively, $\cos2\phi_T$ azimuthal asymmetry decreases as $K_\perp$  increases.

Fig.~\ref{fig:ub_vzh} shows the $z_h$ variation for two different virtualities of the photon of the upper bound of the $\cos2\phi_T$ (left panel) and $\cos2(\phi_T-\phi_{\perp})$ (right panel) azimuthal asymmetries. The upper bounds increase as the virtuality of the photon increases, and both azimuthal asymmetries show a maximum at $z_h \approx 0.3$. The upper bound of  $\cos2(\phi_T-\phi_{\perp})$ azimuthal asymmetry becomes zero and then changes sign at higher values of $z_h$. This is due to a change in the sign of the coefficient $\mathcal{B}_1$ in the numerator.
In Fig.~\ref{fig:ub_vy} the upper bounds for ${\cos2\phi_T}$ (left panel) and ${\cos2(\phi_{T}-\phi_\perp)}$ (right panel) are plotted as a function of $y$. 
As $y$ increases, the magnitude of ${\cos2\phi_T}$ azimuthal asymmetry decreases and reaches its minimum at $y=1$, due to the vanishing of the coefficient $\mathcal{B}_0$ at $y=1$.
For ${\cos2(\phi_{T}-\phi_\perp)}$, the coefficient $\mathcal{B}_2$ contributes, which involves both longitudinal and transverse polarization of the photon. At $y=1$, only the contribution from transverse photons leads to a larger asymmetry.  
The magnitude of ${\cos2\phi_{T}}$ azimuthal asymmetry increases as the virtuality of the photon increases from $Q^2=10$ $\mathrm{GeV^{2}}$ to  $Q^2=20$ $\mathrm{GeV^{2}}$.
In contrast, the magnitude of the upper bound of ${\cos2(\phi_{T}-\phi_\perp)}$ is larger for $Q^2=10$ $\mathrm{GeV^{2}}$ compared to $Q^2=20$ $\mathrm{GeV^{2}}$ for low values of $y$.
The upper bound of  $\cos2(\phi_{T}-\phi_\perp)$ azimuthal asymmetry becomes zero and then changes sign, because the numerator switches the sign from positive to negative. The value of $y$ where this happens depends on the photon virtuality.

\begin{figure}[H]
	\begin{center} 
		\begin{subfigure}{0.49\textwidth}
		\includegraphics[height=7cm,width=8.5cm]{Fig5a.pdf}
			\caption{}
		\end{subfigure}
	    \begin{subfigure}{0.49\textwidth}
	    \includegraphics[height=7cm,width=8.5cm]{Fig5b.pdf}
	        \caption{}
	    \end{subfigure}
     \end{center}
     \caption{\label{fig:ub_vzh} Upper bound for the $A^{\cos2\phi_T}$ (left panel) and $A^{\cos2(\phi_{T}-\phi_\perp)}$ (right panel) azimuthal asymmetries in $e+p\rightarrow e +D+\mathrm{jet}+X$ process as function of $z_h$ at {EIC $\sqrt{s}=45$ GeV for} fixed values of $y=0.1$, $K_\perp=2$ GeV {and $Q^2=10$, 20 GeV$^2$}. The {kinematical variables} $z$ {and $q_T$ are integrated from [0,1]}.}
\end{figure}


\subsection{Gaussian parameterization}
\begin{figure}[H]
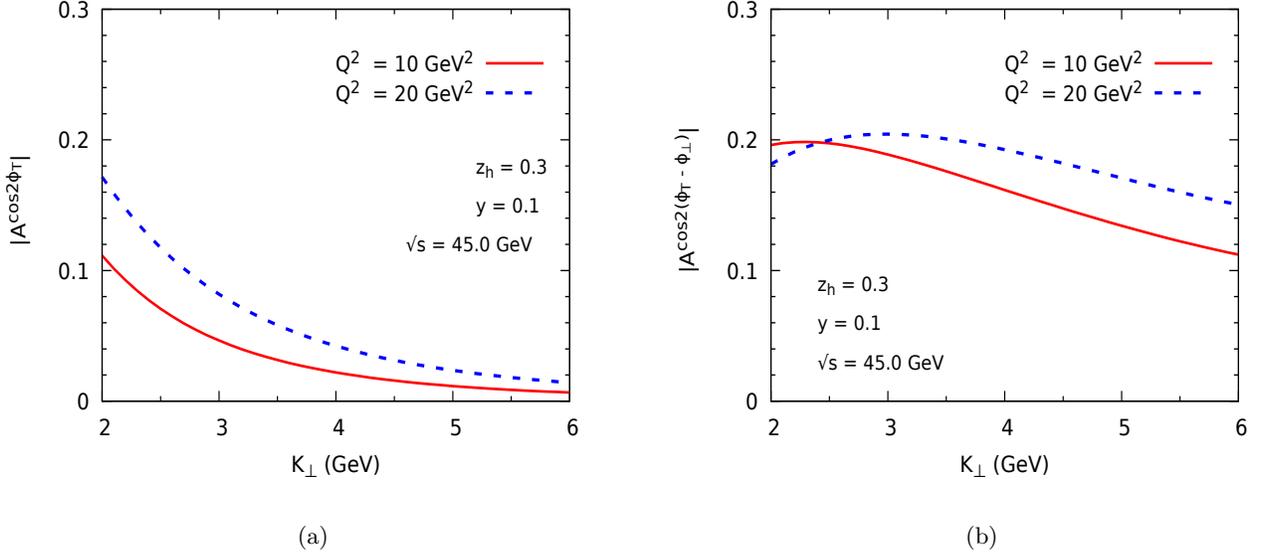

	\begin{center} 
		\begin{subfigure}{0.49\textwidth}
		\includegraphics[height=7cm,width=8.5cm]{Fig7a.pdf}
			\caption{}
		\end{subfigure}
	    \begin{subfigure}{0.49\textwidth}
	    \includegraphics[height=7cm,width=8.5cm]{Fig7b.pdf}
	        \caption{}
	    \end{subfigure}
	\end{center}
\caption{\label{fig:ubgau_vKp} Absolute values of $\cos 2\phi_{T}$ (left panel) and $\cos 2(\phi_{T}-\phi_{\perp})$ (right panel) azimuthal asymmetries in $e+p\rightarrow e +D+\mathrm{jet}+X$ process as function of $K_\perp$ for $\sqrt{s}=45$ GeV at fixed values of $y=0.1$, $z_h=0.3$  for two values of $Q^2=10, 20$ $\mathrm{GeV}^{2}$. The $z$ and $q_T$ are integrated over $0 < z<1,~0<q_T < 1 ~ \mathrm{GeV}$.}
\end{figure}

\begin{figure}[H]
	\begin{center} 
		\begin{subfigure}{0.49\textwidth}
		\includegraphics[height=7cm,width=8.5cm]
         {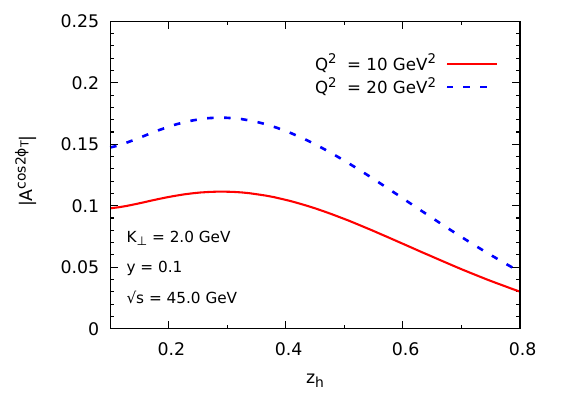}
			\caption{}
		\end{subfigure}
	    \begin{subfigure}{0.49\textwidth}
	    \includegraphics[height=7cm,width=8.5cm]{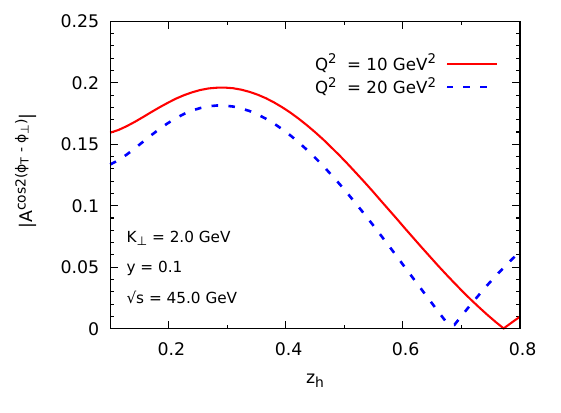}
	        \caption{}
	    \end{subfigure}
	\end{center}
\caption{\label{fig:ubgau_vzh} Absolute values of $\cos 2\phi_{T}$ (left panel) and $\cos 2(\phi_{T}-\phi_{\perp})$ (right panel) azimuthal asymmetries in $e+p\rightarrow e+D+\mathrm{jet}+X$ process as function of $z_h$ for $\sqrt{s}=45$ GeV at fixed values of $y=0.1$, $K_{\perp}=2~ \mathrm{GeV}$  for two values of $Q^2=10, 20$ $\mathrm{GeV}^{2}$. The $z$ and $q_T$ are integrated over $0 < z<1,~0<q_T < 1 ~ \mathrm{GeV}$.}
\end{figure}

\begin{figure}[H]
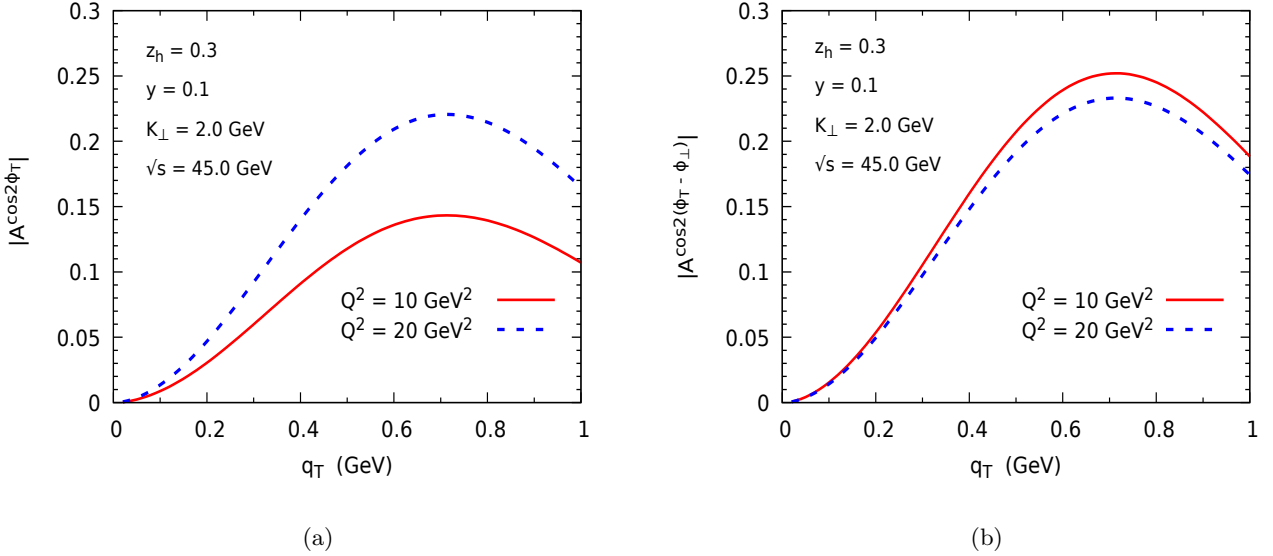

	\begin{center} 
		\begin{subfigure}{0.49\textwidth}
		\includegraphics[height=7cm,width=8.5cm]{Fig10a.pdf}
			\caption{}
		\end{subfigure}
	    \begin{subfigure}{0.49\textwidth}
	    \includegraphics[height=7cm,width=8.5cm]{Fig10b.pdf}
	        \caption{}
	    \end{subfigure}
	\end{center}
	\caption{\label{fig:gauss_vqt_fix} Absolute values of $\cos 2\phi_{T}$ (left panel) and $\cos 2(\phi_{T}-\phi_{\perp})$ (right panel) azimuthal asymmetries in $e+p\rightarrow e +D+\mathrm{jet} +X$ process as function of $q_T$ for $\sqrt{s}=45$ GeV at fixed values of  $K_\perp=2~\text{GeV},~y=0.1~\&~ z_h=0.3$ for two values of $Q^2$. The $z$ is integrated over $0 < z < 1$. }
\end{figure}

In this section, we present the numerical results obtained by parameterizing the gluon TMDs using the Gaussian parameterization  {provided} in Eqs.~(\ref{eq:gauss_f1}) and (\ref{eq:gauss_h1p}). In  Fig.~\ref{fig:ubgau_vKp}-\ref{fig:gauss_vqt_fix}, $\cos2\phi_T$ (left panel) and $\cos2(\phi_T-\phi_\perp)$ (right panel) azimuthal asymmetries are shown as  functions of $K_\perp$, $z_h$, $y$ and $q_T$, {respectively} at $\sqrt{s}=45$ GeV. In this plots, the kinematical variables are chosen to maximize the asymmetry.  
In  Fig.~\ref{fig:ubgau_vKp}, we compare the asymmetries for two different virtualities of the  photon, $Q^2$. From Fig.~\ref{fig:ubgau_vKp}, one can see that ${\cos2\phi_T}$ 
asymmetry is higher  for higher  $Q^{2}$ value, but in the low $K_{\perp}$ region, 
${\cos2(\phi_{T}-\phi_\perp)}$ asymmetry is larger for lower value of $Q^2$. Moreover, the asymmetries decrease as $K_{\perp} $ increases. However, the $\cos2\phi_T$ asymmetry decreases much faster compared to ${\cos2(\phi_{T}-\phi_\perp)}$.
The variation of both the azimuthal asymmetries as a function of $z_h$ is shown in Fig.~\ref{fig:ubgau_vzh}. For both $Q^2$ values, the azimuthal asymmetries are maximum 
at $z_h=0.3$. As shown in ${\cos2\phi_T}$ plot, the  asymmetry {initially} increases with $z_h$, reaches a maximum value, and then decreases.
In the ${\cos2(\phi_{T}-\phi_\perp)}$ plot, the asymmetry increases first and reaches its maximum value. After that, it decreases to zero and then becomes negative  with increasing $z_h$. This qualitative behavior depends on the relative dominance of the term with  transverse polarization of photons (first term of Eq. \eqref{lep:ep1}) and the term with longitudinal polarization (second term of Eq. \eqref{lep:ep1}).  The magnitude of ${\cos2(\phi_{T}-\phi_\perp)}$ vanishes  at $z_h=0.6$ for $Q^2=20$ $\mathrm{GeV}^{2}$ and at $z_h=0.7$ for $Q^2=10$ $\mathrm{GeV}^{2}$.
Unlike the ${\cos2\phi_T}$ asymmetry, the ${\cos2(\phi_{T}-\phi_\perp)}$ asymmetry is larger for a lower value of $Q^2$. The $y$ variation of  ${\cos2\phi_T}$ and ${\cos2(\phi_{T}-\phi_\perp)}$ azimuthal asymmetries is shown in Fig.~\ref{fig:ubgau_vy}. The $\cos2\phi_T$ azimuthal asymmetry decreases monotonically as $y$ increases. On the other hand,  the ${\cos2(\phi_{T}-\phi_\perp)}$ azimuthal asymmetry shows different behavior in the low and high $y$ regions. In the low $y$ region, the ${\cos2(\phi_{T}-\phi_\perp)}$ azimuthal asymmetry shows a similar behavior to the ${\cos2\phi_T}$ azimuthal asymmetry. As $y$ increases, the asymmetry becomes zero ($y\approx 0.7$ for $Q^2=20$ $\mathrm{GeV}^{2}$ and at $y\approx 0.85$ for $Q^2=10$ $\mathrm{GeV}^{2}$) and then becomes negative. As discussed above, this behavior is due to a relative dominance of the contributions from the transversely and longitudinally polarized photon. In the limit $y\rightarrow 1$, the ${\cos2\phi_T}$ azimuthal asymmetry vanishes since the coefficient $\mathcal{B}_0$ as given in Eq.\eqref{eq:B0} vanishes in this limit. This happens because the contribution from the longitudinally polarized photon vanishes.
For ${\cos2\phi_T}$, only the longitudinally polarized photon contributes, whereas for ${\cos2(\phi_{T}-\phi_\perp)}$, both the longitudinally and transversely polarized photon contribute. As $y\rightarrow 1$, the ratio of $\mathcal{B}_2/\mathcal{A}_0$ which probes the ${\cos2(\phi_{T}-\phi_\perp)}$ asymmetry comes only from transversely polarized photons. As seen in the $z_h$ dependent plots, ${\cos2(\phi_{T}-\phi_\perp)}$ asymmetry is larger for lower value of $Q^2$, 
whereas ${\cos2\phi_T}$ asymmetry is larger for higher value of $Q^2$. 

In Fig.~\ref{fig:gauss_vqt_fix}, the $q_T$ variation is shown and is Gaussian in nature due to the parameterization of TMDs.
Both the asymmetries show a maximum at $q_T \approx 0.7$ $\mathrm{GeV}$. The position of the maximum is independent of $Q^2$, however, the magnitude depends on $Q^2$.
The magnitude of ${\cos2\phi_T}$ increases as the virtuality of the photon increases, whereas the magnitude of ${\cos2(\phi_{T}-\phi_\perp)}$ decreases as $Q^2$ increases. Overall, from these plots one can see that the asymmetries are quite sizable in the kinematics of EIC, reaching about $20-25 \%$ in certain regions.  

\begin{figure}[H]
	\begin{center} 
		\begin{subfigure}{0.49\textwidth}
		\includegraphics[height=7cm,width=8.5cm]
           {Fig9a.pdf}
			\caption{}
		\end{subfigure}
	    \begin{subfigure}{0.49\textwidth}
	    \includegraphics[height=7cm,width=8.5cm]{Fig9b.pdf}
	        \caption{}
	    \end{subfigure}
	\end{center}
\caption{\label{fig:ubgau_vy} Absolute values of $\cos 2\phi_{T}$ (left panel) and $\cos 2(\phi_{T}-\phi_{\perp})$ (right panel) azimuthal asymmetries in $e+p\rightarrow e +D+\mathrm{jet}+X$ process as function of $y$ for $\sqrt{s}=45$ GeV at fixed values of $z_{h}=0.3$, $K_{\perp}=2$ GeV  for two values of $Q^2=10, 20$ $\mathrm{GeV}^{2}$. The $z$ and $q_T$ are integrated over $0 < z<1,~0<q_T < 1 ~ \mathrm{GeV}$.}
\end{figure}

\begin{figure}[H]
	\begin{center} 
	\begin{subfigure}{0.49\textwidth}
	\includegraphics[height=7cm,width=8.5cm]{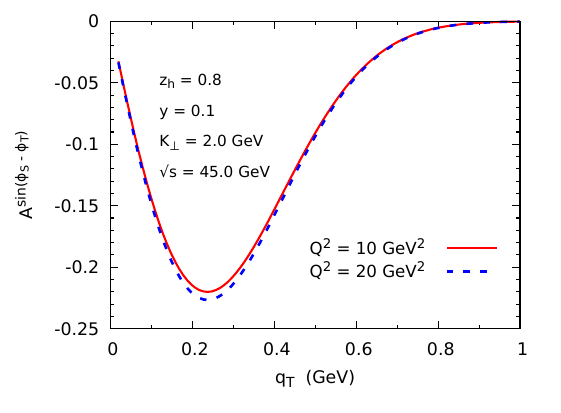}
		\caption{}
	\end{subfigure}
	    \begin{subfigure}{0.49\textwidth}
	    \includegraphics[height=7cm,width=8.5cm]{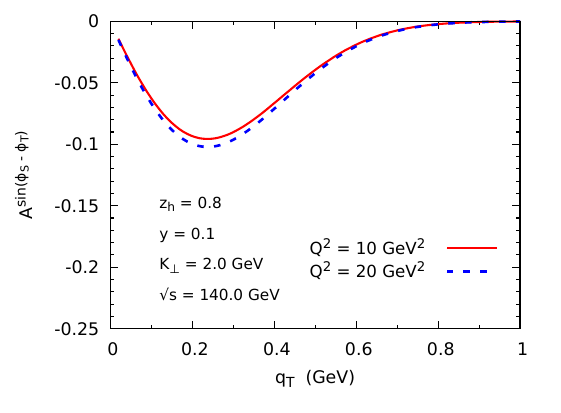}
	        \caption{}
	    \end{subfigure}
	\end{center}
\caption{\label{fig:sphis-pht_vqT} Sivers asymmetry in $e+p^{\uparrow}\rightarrow e +D+\mathrm{jet}+X$ process as function of $q_T$ at fixed values of $y=0.1$, $z_h=0.8$ and $K_{\perp}=2$ GeV  for $\sqrt{s}=45~ \mathrm{GeV}$ (left panel) and  $\sqrt{s}=140 ~\mathrm{GeV}$ (right panel) for two values of $Q^2$. The $z$ is integrated over $0 < z <1$. }
\end{figure}

\begin{figure}[H]
	\begin{center} 
	\begin{subfigure}{0.49\textwidth}
\includegraphics[height=7cm,width=8.5cm]{Fig12a.pdf}
  \caption{}
	\end{subfigure}
	    \begin{subfigure}{0.49\textwidth}
      \includegraphics[height=7cm,width=8.5cm]{Fig12b.pdf}
	        \caption{}
	    \end{subfigure}
	\end{center}
\caption{\label{fig:sphis-pht_vy} Sivers asymmetry in $e+p^{\uparrow}\rightarrow e +D+\mathrm{jet}+X$ process as function of $y$ at fixed values of $K_\perp=2.0 ~\mathrm{GeV}$, $z_h=0.8$  for $\sqrt{s}=45~ \mathrm{GeV}$ (left panel) and  $\sqrt{s}=140 ~\mathrm{GeV}$ (right panel) for two values of $Q^2$. The $z$ and $q_T$ are integrated over $0 < z<1,~0<q_T < 1 ~ \mathrm{GeV}$. }
\end{figure}

\begin{figure}[H]
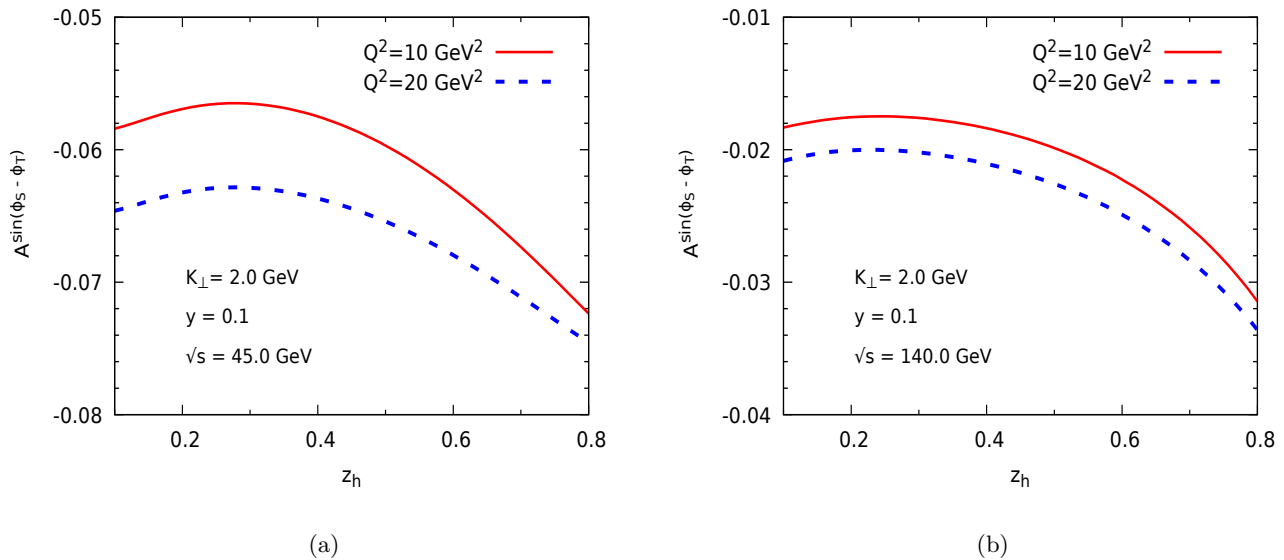

	\begin{center} 
	\begin{subfigure}{0.49\textwidth}
\includegraphics[height=7cm,width=8.5cm]{Fig13a.pdf}
  \caption{}
	\end{subfigure}
	    \begin{subfigure}{0.49\textwidth}
    \includegraphics[height=7cm,width=8.5cm]{Fig13b.pdf}
	        \caption{}
	    \end{subfigure}
	\end{center}
\caption{\label{fig:sphis-pht_vzh} Sivers asymmetry in $e+p^{\uparrow}\rightarrow e +D+\mathrm{jet}+X$ process as function of $z_h$ at fixed values of $K_\perp=2.0 ~\mathrm{GeV}$, $y=0.1$  for $\sqrt{s}=45~ \mathrm{GeV}$ (left panel) and  $\sqrt{s}=140 ~\mathrm{GeV}$ (right panel) for two values of $Q^2$. The $z$ and $q_T$ are integrated over $0 < z<1,~0<q_T < 1 ~ \mathrm{GeV}$. }
\end{figure}

In  Fig.~\ref{fig:sphis-pht_vqT}, the Sivers asymmetry is shown at two different cm energies, $\sqrt{s}=45~ \mathrm{GeV}$  and $\sqrt{s}=140~ \mathrm{GeV}$, respectively,  for two different virtualities of the photon, and a Gaussian parameterization for the gluon Sivers function. It is seen from the plot that the Sivers asymmetry is negative. The asymmetry is quite sizable in our kinematics, for $\sqrt{s}=45~ \mathrm{GeV}$ the peak is about $23 \%$ whereas, for the higher energy, the peak is about $10 \%$. The position of the peak is independent of the  cm energy and is at $q_T\approx 0.2 ~\mathrm{GeV}$ for both energies.
From Fig.~\ref{fig:sphis-pht_vqT}, one can see that the Sivers asymmetry is high for a low cm energy $i.e.$ for $\sqrt{s}=45~ \mathrm{GeV}$. This is due to the $\mathcal{N}_{g}(x)$ term of the Sivers function given in the Gaussian parameterization model. The $\mathcal{N}_{g}(x)$ term inversely depends on the cm energy through the $x$ defined in Eq.\eqref{dfun2}. As cm energy  increases the $x$ and $\mathcal{N}_{g}(x)$ decreases, which results in the decrease of the Sivers asymmetry.
Additionally, one can see that the  asymmetry does not depend that much on  $Q^2$.\par

Fig.~\ref{fig:sphis-pht_vy} and Fig.~\ref{fig:sphis-pht_vzh} show the variation of Sivers asymmetry as a function of the inelasticity $y$ 
 and momentum fraction $z_h$ for two different values of photon virtuality $(Q^2)$  at two different cm energies ($\sqrt{s}=45$ and $140$ GeV). Here, the plots show the negative asymmetry. In Fig.~\ref{fig:sphis-pht_vy},  the magnitude of Sivers asymmetry decreases as the value of $y$ increases.
 For $y$ variation, the contribution from transversely polarized photons is significantly larger, approximately one order of magnitude higher, when compared to the contribution from longitudinally polarized photons.
  This is observed throughout the range of $y$ values, the transversely polarized photon contribution decreases as $y$ increases. Notably, at $y=1$, the contribution mainly comes from the transversely polarized photon, resulting in a non-vanishing asymmetry.
 The asymmetry does not depend significantly on the photon virtuality.
In  Fig.~\ref{fig:sphis-pht_vzh}, both transversely polarized and longitudinally polarized photons contribute, with the transversely polarized photon making the dominant contribution. The magnitude of the Sivers asymmetry is maximum at $z_h=0.8$ and it decreases for lower values of $z_h$. Furthermore, it is observed that the asymmetry is large for lower values of $Q^2$.

\section{Conclusion}
In this article, we have investigated the azimuthal asymmetries in $D$-meson and jet production in the process of electron-proton collision in the kinematics of the future Electron-Ion Collider.
We have considered the kinematical condition where the final particles $D$-meson and jet are almost back-to-back in the plane perpendicular to the direction of the incoming proton and the photon exchanged in the process and we used the TMD factorization formalism. The $D$-meson is produced from the fragmented charm quark in the photon-gluon fusion subprocess. We presented numerical estimates of the azimuthal asymmetries for this process; we calculated the model-independent upper bounds, as well as estimated the asymmetries using a widely used Gaussian parameterization of the TMDs. The  $\cos2\phi_{T}$  and $\cos2(\phi_{T}-\phi_{\perp})$ azimuthal modulations  in the unpolarized cross-section allow us to probe the linearly polarized gluon TMD. Our numerical estimates of the asymmetries in the kinematics of EIC show that they are sizable, and can be as large  as $20 \%$ in certain kinematical regions. The $\cos2(\phi_{T}-\phi_{\perp})$ shows a sign change due to competing contributions from transverse and longitudinally polarized virtual photons. When the proton is transversely polarized we estimated the Sivers azimuthal modulation, $\sin(\phi_{S}-\phi_{T})$, which could  probe the gluon Sivers TMD. We obtained a sizable Sivers asymmetry in the kinematics considered, which will be accessible at the EIC. Our calculations show that $D$-meson and jet production at the EIC could be a useful process to probe the gluon TMDs.

\section{Acknowledgement}
AM would like to thank SERB MATRICS (File no. MTR/2021/000103) for support. KB acknowledges support from CFNS through the joint IITB-CFNS postdoc position.  
\appendix
\section{Amplitude modulations }\label{appen}
We redefine the partonic Mandelstam variables as the following
\begin{eqnarray*}
 s&=&Q^2\left(\frac{x-\xB}{\xB}\right)\,,\\
 u&=&m_c^2-z_1x\frac{Q^2}{\xB}\,,\\
t&=&m_c^2\left(\frac{z_1-1}{z_1}\right)+Q^2(z_1-1)-\frac{K^2_{\perp}}{z_1}\,.
\end{eqnarray*}
The amplitude modulations are listed here
\begin{align}
 \mathcal{A}_0&=-\frac{ 1}{  Q^2 \left(Q^2+s\right)^2 \left(m_c^2-u\right)^2
   \left(m_c^2-t\right)^2}\Bigg\{(1+(1-y)^2 ) \Big[3 Q^{12}+4 Q^{10} (5 s+3 (t+u))+\nonumber\\
   & Q^8 \left(53 s^2+60 s (t+u)+4 \left(t^2+6 t u+u^2\right)\right)+4 Q^6 \big(18 s^3+30
   s^2 (t+u)+2 s (3 t+u) (t+3 u)-\nonumber\\
   &(t-u)^2 (t+u)\big)+Q^4 \left(53 s^4+120 s^3 (t+u)+48 s^2 (t+u)^2-4 s (t-u)^2 (t+u)+(t-u)^4\right)\nonumber\\
   &+4 Q^2 s^2 \left(5 s^3+15 s^2 (t+u)+2 s \left(5 t^2+6 t u+5
   u^2\right)+(t-u)^2 (t+u)\right)+s^2 \big(3 s^4+12 s^3 (t+u)+\nonumber\\
   &4 s^2 \left(3 t^2+2 t u+3 u^2\right)+4 s (t-u)^2 (t+u)+(t-u)^4\big)\Big]+8(1-y)Q^2\left(m_c^2-u\right)
   \left(m_c^2-t\right)\Big[  2 Q^6+\nonumber\\
   &Q^4 (5 s+2 (t+u))+4 Q^2 s (s+t+u)+s \left(s^2+2 s (t+u)+(t-u)^2\right)\Big]
   \Bigg\}
\end{align}

\begin{align}
 \mathcal{A}_1&=\frac{ 8\sqrt{1-y} (y-2)P_{hT}}{ z Q \left(Q^2+s\right) \left(m_c^2-u\right)^2
   \left(m_c^2-t\right)^2}(t-u) \Big[  5 Q^6+Q^4 (13 s+4 (t+u))+Q^2 \big(11 s^2+8 s (t+u)-\nonumber\\
   &(t-u)^2\big)+s \left(3 s^2+4 s
   (t+u)+(t-u)^2\right) \Big]
\end{align}

\begin{align}
 \mathcal{A}_2&=\frac{ 4 (y-1)}{ Q^2 \left(Q^2+s\right)^2 \left(m_c^2-u\right)^2
   \left(m_c^2-t\right)^2} \Big[ \big(2 Q^6+Q^4 (5 s+2 (t+u))+4 Q^2 s (s+t+u)+\nonumber\\
   &s \left(s^2+2 s (t+u)+(t-u)^2\right)\big) \left(3 Q^6+2 Q^4 (4 s+t+u)+Q^2 \left(7
   s^2+4 s (t+u)-(t-u)^2\right)+2 s^2 (s+t+u)\right)\Big]
 \end{align}

\begin{align}\label{eq:B0}
 \mathcal{B}_0&=\frac{ y-1}{ Q^2 \left(Q^2+s\right)^2 \left(m_c^2-u\right)^2
   \left(m_c^2-t\right)^2}\left(3 Q^6+2 Q^4 (4 s+t+u)+Q^2 \left(7 s^2+4 s (t+u)-(t-u)^2\right)+2 s^2 (s+t+u)\right)^2
 \end{align}
 \begin{align}
     \mathcal{B}_1&=\frac{2\sqrt{1-y}(y-2)(t-u)P_{hT} }{z  Q \left(Q^2+s\right) \left(m_c^2-u\right)^2
   \left(m_c^2-t\right)^2}\left( \left(3 Q^6+2 Q^4 (4 s+t+u)+Q^2 \left(7 s^2+4 s (t+u)-(t-u)^2\right)+2 s^2 (s+t+u)\right)\right)
 \end{align}
 
  \begin{align}
    \mathcal{B}_2&=-\frac{ 1}{  Q^2 \left(Q^2+s\right)^2 \left(m_c^2-u\right)^2
   \left(m_c^2-t\right)^2} \Bigg\{(1+(1-y)^2 )
   \big(2 Q^6+Q^4 (5 s+2 (t+u))+4 Q^2 s (s+t+u)+\nonumber\\
   &s \left(s^2+2 s (t+u)+(t-u)^2\right)\big) \left(Q^6+2 Q^4 (2 s+t+u)+Q^2 \left(5 s^2+4 s (t+u)-(t-u)^2\right)+2 s^2 (s+t+u)\right)+\nonumber\\
  & 4(1-y) \left(m_c^2-u\right)
   \left(m_c^2-t\right)\left(2 Q^6+Q^4 (5 s+2 (t+u))+4 Q^2 s (s+t+u)+s \left(s^2+2 s (t+u)+(t-u)^2\right)\right)
   \Bigg\}
 \end{align}
 
 \begin{align}
    \mathcal{B}_3&=\frac{2\sqrt{1-y}(y-2)(t-u)P_{hT} }{z  Q \left(Q^2+s\right) \left(m_c^2-u\right)^2
   \left(m_c^2-t\right)^2} \left(\left(2 Q^6+Q^4 (5 s+2 (t+u))+4 Q^2 s
   (s+t+u)+s \left(2 u (s-t)+(s+t)^2+u^2\right)\right)\right)
 \end{align}

 \begin{align}
     \mathcal{B}_4&=\frac{ y-1}{  Q^2 \left(Q^2+s\right)^2 \left(m_c^2-u\right)^2
   \left(m_c^2-t\right)^2} \left(2 Q^6+Q^4 (5 s+2 (t+u))+4 Q^2 s (s+t+u)+s \left(2 u (s-t)+(s+t)^2+u^2\right)\right)^2
 \end{align}
\bibliographystyle{apsrev}
\bibliography{dmeson_ref}

\end{document}